\documentclass[preprint,12pt]{elsarticle}

\usepackage{lineno,hyperref}
\usepackage{amssymb}
\usepackage{braket}
\usepackage[super,compress]{cite}
\usepackage{epsfig}
\usepackage{epstopdf}
\usepackage{multirow}
\usepackage{booktabs}
\journal{New Astronomy}

\begin{document}

\begin{frontmatter}

\title{Neutrino cooling rates due to nickel isotopes for presupernova
evolution of massive stars}

\author{Jameel-Un Nabi}
\address{Faculty of Engineering Sciences,\\GIK
Institute of Engineering Sciences and Technology,\\ Topi 23640, Khyber
Pakhtunkhwa, Pakistan\\
{jameel@giki.edu.pk}}

\author{Ramoona Shehzadi\footnote{Corresponding Author}}
\address{Department of Physics,\\
University of the Punjab, 54590 Lahore, Pakistan\\
ramoona.physics@pu.edu.pk}

\author{Muhammad Majid}
\address{Faculty of Engineering Sciences,\\GIK
Institute of Engineering Sciences and Technology,\\ Topi 23640, Khyber
Pakhtunkhwa, Pakistan\\
majid.phys@gmail.com}

\begin{abstract}
As per simulation studies, the weak reaction rates on nickel
isotopes play a substantial role in affecting the ratio of
electron-to-baryon content of stellar interior during the late
stages of core evolution. (Anti)neutrinos are produced in weak-decay
processes, and escape from the stellar content having densities less
than 10$^{11}$ g$\;$cm$^{-3}$. They take away energy and reduce the
stellar core entropy. In this paper we report on the microscopic
calculation of neutrino and antineutrino cooling rates due to weak
rates on nickel isotopes in mass range $56\leq A \leq71$. The
calculations are accomplished by employing the deformed pn-QRPA
model. Recent studies on GT strength properties of nickel isotopes
show that the deformed pn-QRPA model well explained the experimental
charge-changing transitions. Our calculated beta decay half-lives
for selected nickel isotopes are in excellent comparison with
experimental data. The (anti)neutrino cooling rates are determined
over temperatures in the range of 0.01$\times$10$^{9}$ --
30$\times$10$^{9}$K and densities in the range of 10 --
10$^{11}$ g$\;$cm$^{-3}$ domain.  The computed rates are compared with
previous theoretical calculations. For neutron rich nuclide, at high
temperatures, our computed cooling rates are enhanced as compared to
previous calculations.
\end{abstract}


\end{frontmatter}

%\linenumbers

%%\tableofcontents
%
\section{Introduction}
Massive stars ($> 8\;$M$_\odot$) play a fundamental role in the
progression of the universe. Their life begins by combustion of
hydrogen in their core under the conditions of hydrostatic burning.
The exhaustion of hydrogen induces the burning of next fuel, helium,
until it finally results into the formation of an iron core, which
halts the process of nuclear energy generation. When core's mass
surpasses the Chandrasekhar limit of about 1.4$\;$M$_\odot$, the
degeneracy pressure of electrons becomes insufficient to counter the
force of gravity and renders the core instable. This may lead to a
process known as core-collapse supernova (type II supernova) which
marks the death of a massive star.

Supernova explosions are considered as major means by which elements
synthesized in stars are injected into the interstellar medium and
recycled into formation of new generations of stars. Therefore, they
are important in controlling much of the chemical enrichment of
galaxies. Apart from the dissemination of the nuclei produced in the
stellar evolution, the explosion also synthesizes elements heavier
than iron via r-process nucleosynthesis~\cite{Cow04}. Therefore, a
broader comprehension into the dynamics of explosion is necessary
for analyzing the nucleosynthesis problem and fate of star's life.

For almost half a century, understanding the mechanism of supernova
explosion has been in forefront of research in the field of
astronomy and astrophysics. The supernova explosion still largely
remains a mystery. The complexity of the explosion mechanism calls
for a complete analysis of the physical processes.
Weak-interaction-mediated processes are considered important for an
improved understanding of the presupernova phases of stellar
evolution and hence their extensive consideration becomes necessary.
Weak processes, e.g., nuclear $\beta$-decay and electron capture,
alter the overall lepton-to-baryon ratio (Y$_e$) of the presupernova
star~\cite{Lang03} as well as its core entropy. The number or
quantity of electrons present for pressure support are decreased via
electron capture processes whereas $\beta$-decay acts oppositely.
Thus, these nuclear reactions are vital for a better understanding
of  the collapse dynamics.

Weak interactions also result in copious production of
(anti)neutrinos which for stellar densities, up to $\rho \le
10^{11}\;$g$\;$cm$^{-3}$, escape the core and assist in the cooling of
core by taking away its energy and
entropy~\cite{Heger01, Janka07}. At higher core densities
($\sim 10^{12}\;$g$\;$cm$^{-3}$), where neutrino diffusion time
exceeds the time required to complete the collapse~\cite{Bethe90},
the neutrinos are trapped by elastic scattering on nucleons. When
the inner core reaches nuclear densities ($\rho \sim
10^{14}\;$g$\;$cm$^{-3}$), the collapse halts and produces a shock
wave with energy. The outward progress of the shock wave is stalled
by energy losses from the disintegration of heavy nuclei across the
shock and through neutrino emission. This standing shock is thought
to be energized by neutrino-heating mechanism, as suggested by
Wilson~\cite{Wilson85} and Bethe $\&$ Wilson~\cite{Bethe85}.
However, even after 3 decades, the simulation studies of the
neutrino-heating mechanism has not confirmed it as the trigger of
explosion. Now, there is a large agreement that one dimensional
models generally fail to convert the collapse into an explosion
except for the low mass progenitor
stars~\cite{Bruen01, Janka08}. This has lead a shift towards
two- and three-dimensional (3D) models where multidimensional
effects, such as rotation, convection, turbulence and instabilities
are considered (see e.g.,~\cite{Fryer00, Melson15}). Many of
these works have demonstrated that multidimensional effects are
crucial and generally aid an explosion.

Neutrinos from core-collapse supernova play an indispensable part to
understand the microphysics of the supernova. They act as a major
source of transporting and removing energy from collapsing stars.
They can provide information about various events from collapse to
explosion, e.g., the neutronization, the bounce and shock formation,
propagation and stagnation of the shock and the phase of neutrino
cooling. Neutrino energy loss or cooling rates calculated over a
broad domain of stellar temperature and density scales are essential
input parameters for core-collapse simulation of massive
stars~\cite{Strother09}. A reliable and microscopic computation of
these rates is necessary to execute a thorough investigation of the last
phases of stellar development.

Nuclear weak interactions during final stages of the evolvement of a
high mass star are known to be governed by Gamow-Teller (GT)
charge-changing transitions~\cite{Bethe79}. In particular GT
transition strengths for $fp$-/$fpg$-shell nuclide are of
fundamental significance for core-collapse supernova
physics~\cite{FFN} and play a decisive
character in many astrophysical matters, involving stellar evolution
and the related nucleosynthesis problem. At the initial stages of
the collapse, when the electron chemical potential, $\mu_{e}$, is
more or less same as the nuclear Q-values, the detailed calculation
of GT strengths is needed for the relevant weak-interaction rates.
Whereas at a later stage, for higher stellar densities (where
$\mu_{e} > Q$), the electron capture rates largely depend on the
total GT strength. A precise determination of the GT transitions is
a complex problem, as the weak-interaction-mediated processes in the
massive stars occur at relatively high temperatures and GT
transitions from thermally populated excited states can contribute
significantly and hence must be taken into account.

The first considerable attempt to measure the astrophysical weak
rates over broad dimensions of temperature and density was
accomplished by Fuller, Fowler and Newman
(FFN)~\cite{FFN}. They measured not only the
lepton capture and emission rates but also the (anti)neutrino energy
loss/cooling rates of 226 different nuclide having mass in the range
21-60. They employed zeroth-order shell model to estimate the
excitation energies and GT strength distributions by incorporating
measured data present at that time. Later, the FFN work was expanded
for more heavy nuclei with $A > 60$ by Aufderheide and
collaborators~\cite{Aufder94}. They adopted a similar formalism as
used by FFN. In their studies, the quenching of the charge-changing transitions strength was
considered by minimizing the zeroth-order shell model calculation
for the Gamow-Teller resonance contribution by a typical factor of
2. Results of the charge-changing reactions such as, (n,p) and (p,n)
measurements~\cite{Rapaport83, Kateb94}, have exposed the
misplacing of the charge-changing strength centroid used in the calculations of
Aufderheide and FFN. This energized theoretical attempts for
improvements in the weak rates by considering their microscopic
calculations. Among several theories, the deformed
pn-QRPA~\cite{Nabi99} and the large-scale shell model
(LSSM)~\cite{Lang00} are the most effective theories which have been
applied largely for the microscopic computation of astronomical
weak reaction rates.

The deformed pn-QRPA model has two main advantages over the previous
ways of estimation of weak reaction rates. Firstly, in this model
one can use a huge model space up to $7\hbar\omega$ shells for
calculation. This space suffices for the calculation of strength
functions for any arbitrary heavy nucleus. Secondly, this model does
not utilize the Brink-Axel hypothesis \cite{BAH} (as employed in the
computations of LSSM and FFN) to evaluate GT transitions from parent
excited states. It, on the contrary, permits a microscopic state by
sate estimation of GT charge-changing transitions from ground and
excited levels. This feature of the pn-QRPA model makes it more
reliable for the estimation of astronomical weak rates and thus is
being successfully used in recent calculations. Nabi and
Klapdor-Kleingrothaus \cite{Nabi99} were the first who used this
model for the computation of weak-decay rates over a broad
dimensions of density and temperature for 709 nuclei ($18 \le A \le
100$) in the stellar matter. They have performed calculation of lepton capture
and decay rates, the associated energy loss/cooling
weak-rates, probabilities and energy rates of $\beta$-delayed particle
emissions, and gamma heating rates in stellar matter~\cite{Nabi99a,
Nabi04}. The calculations were later further refined, on
case-to-case basis (e.g., Refs.~\cite{Nabi07, Nabi08a}), using more
proficient algorithms, embodiment of latest experimental results and
improvement of the model parameters. A discussion on the accuracy
and reliability of the pn-QRPA theory may be found in~\cite{Nabi04}.

Nickel isotopes are abundant in the presupernova environment and are
considered to play a substantial part in the evolutionary process of
high mass stars. The simulation studies of Aufderheide et
al.,~\cite{Aufder94} and of Heger et al.~\cite{Heger01a} of the late
phases of development of massive stars show that the electron
captures (EC) on $^{56-61, 63-65, 67-69}$Ni and $\beta$-decay rates
of $^{63,65-69,71}$Ni isotopes are significantly important in
affecting $Y_{e}$ in stellar environment. The results of GT strength
distribution and the EC rates due to $^{56}$Ni, estimated using the
pn-QRPA model were first reported and compared with previous
measurements in Ref.~\cite{Nabi05}. The possible employment of these
calculated rates in the stellar environment was also discussed
there. The calculations were later improved and extended to heavier
nickel isotopes (with the mass range $A$ = 57-65)~\cite{Nabi12}.
Recently, the GT$_\pm$ transitions properties and weak-decay rates,
on Ni isotopes were calculated in Refs.~\cite{Nab17a, sad18}. In the
present work, the deformed pn-QRPA model is used to compute the
(anti)neutrino cooling rates of nickel isotopic chain having mass
number between 56 and 71. Our results have also been compared with
previous results computed by FFN, LSSM and by Pruet and Fuller
(PF)~\cite{Pruet03}.

The paper is organized as follows.  In the next section, the
theoretical formalism used in the computation of GT strength and the
related (anti)neutrino cooling weak-rates has been presented. Our
results and their comparison with other model calculations are given
in Section 3. In the last section, we summarize our work and present
the conclusions.
\section{Formalism}
The deformed pn-QRPA model was considered for the theoretical
calculation of (anti)neutrino cooling rates of nickel (Ni)
isotopes. The Hamiltonian (H) selected for this model was considered
as
\begin{equation}
H = H^{SP} + V^{Pairing} + V_{GT}^{PH} + V_{GT}^{PP}.
\end{equation}
where $H^{SP}$ represents the Hamiltonian of single particle ($SP$),
$V^{pairing}$ represents the pairing force (treated in the BCS
approximation), $V_{GT}^{PH}$ and $V_{GT}^{PP}$ are the
particle-hole ($PH$) and particle-particle ($PP$) Gamow-Teller (GT)
potentials (forces), respectively. Nilsson model was used for the
estimation of $SP$ wave functions and energies ~\cite{Nil55}, in
which the nuclear deformation ($\beta_{2}$) was considered. The $PH$
and $PP$ interaction constants were characterized by $\chi$ and
$\kappa$, respectively. These parameters were selected in order to
reproduce the available experimental half-lives and satisfy the
Ikeda sum rule (ISR) \cite{Isr63}. In this work, we took $\chi$
equal to 4.2/A (MeV), displaying a $1/A$ dependence \cite{Hom96,
Nab17, Maj17} and $\kappa$ to be 0.10 MeV. The values of these
parameters are same as considered in the recent calculation of
lepton capture rates for Ni isotopes \cite {Nab17a}. Nilsson
potential parameters were chosen from Ref.~\cite{Nil55}. The
estimated half-lives (T$_{1/2}$) values weakly rely on the $\Delta
_{p}$ and $\Delta _{n}$~\cite{Hir91}. The conventional values
of
\begin{equation}
\Delta_{p} = \Delta_{n} = 12/\sqrt{A} (MeV)
\end{equation}
were taken in this paper. The nuclear deformation $\beta_{2}$ was
determined by using the formula
\begin{equation}
\beta_{2} = \frac{125(Q_2)}{1.44(Z)(A^{2/3})}
\end{equation}
where $Q_2$ denote the electric quadrupole moment chosen from
Ref.~\cite{Mol95}. Q-values of reactions were chosen from Ref.~\cite{Aud12}. The
charge-changing transitions, in pn-QRPA model, are defined in terms
of phonon creation operators. The pn-QRPA phonons are expressed as
\begin{equation}
A_{\omega}^{+}(\mu)=\sum_{pn}(X^{pn}_{\omega}(\mu)a_{p}^{+}a_{\bar{n}}^{+}-Y_{\omega}^{pn}(\mu)a_{n}
a_{\bar{p}}).
\end{equation}
In above equation the summation was taken on all the proton-neutron
pairs having $\mu$ = \textit{m$_{p}$-m$_{n}$} = -1, 0, 1, where
\textit{m$_{p}$}(\textit{m$_{n}$}) represent the angular momentum
third component of proton(neutron). The \textit{a$^{+}_{p(n)}$}
represents the creation operator of a quasi-particle (q.p) state of
proton(neutron), whereas \textit{$\bar{p}$} denotes the time
reversed state of \textit{p}. The ground level of the theory with
respect to the QRPA phonons is expressed as the vacuum,
A$_{\omega}(\mu)|QRPA\rangle$ = 0. The phonon operator
A$_{\omega}^{+}(\mu)$, having excitation energy ($\omega$) and
amplitudes (\textit{X$_{\omega}, Y_{\omega}$}) were achieved by
solving the renowned RPA matrix equation. Detailed solution of RPA
matrix equation can be seen in Refs. \cite{Hir91, Mut89}.

The (anti)neutrinos in stellar environment are produced mainly
through four different weak-interaction processes: by emission of
positron and electron, and due to captures of positron and electron
(from continuum as they are not assumed to be in bound states) . It
is further assumed that during the presupernova evolutionary stages
the (anti)neutrinos produced can pass through the stellar matter and
due to energy transfer the stellar core cools itself. The
(anti)neutrino energy loss weak-rates can be calculated as
\begin{eqnarray}
\lambda ^{^{\nu(\bar{\nu})}} _{nm} &=& \left(\frac{\ln 2}{D} \right)
[f_{nm}^{\nu} (E_{f}, T, \rho)][B(F)_{nm} \nonumber \\
&+&(g_{A}/ g_{V})^{2} B(GT)_{nm}].
\end{eqnarray}
The constant D value was chosen as 6143s \cite{Har09} and
$g_{A}/g_{V}$ was considered to be -1.2694 \cite{Nak10}. B(GT) and B(F) represent the Gamow-Teller and Fermi transition probabilities, respectively and are specified as
\begin{eqnarray}
 B(F)_{nm} &=& \frac{1}{J_{n} + 1} |\langle m ||\sum_{k} \tau^{k}_{\pm} || n \rangle |^{2} \\
 B(GT)_{nm} &=& \frac{1}{J_{n} + 1} |\langle m ||\sum_{k} \tau^{k}_{\pm} \vec{\sigma}^{k} || n \rangle |^{2}
\end{eqnarray}
here $\overrightarrow{\sigma}(k)$ and $\tau_{\pm }^{k}$ show the
spin and the isospin operators, respectively. For the daughter and
parent excited levels construction and estimations of matrix
elements, within the framework of deformed pn-QRPA model, see Ref.
\cite{Nabi99}. The phase space integrals ($f_{nm}^{\nu}$) were taken
over total available energy and were calculated using the following relations.
\begin{equation}
f_{nm}=\int _{1}^{w_{m} }w\sqrt{w^{2} -1}(w_{m} -w)^{2}F(\pm Z,w)(1-
G_{\mp })dw. \label{phdecay}
\end{equation}
(for positron emission lower sign is used while for electron
emission upper sign is used), and the phase space integral for
positron capture (PC) (lower sign) and electron capture (EC) upper
sign is given as
\begin{equation}
f_{nm} = \int _{w_{1} }^{\infty }w\sqrt{w^{2} -1}(w_{m} +w)^{2}
F(\pm Z, w) G_{\mp }dw, \label{phcapture}
\end{equation}
where $w$ represent the electron total kinetic
energy (K.E) plus its rest mass, while the total lepton capture threshold energy
for PC and EC is denoted by $w_{l}$. F($\pm Z$,$w$) represent the
Fermi functions and are determined by using the same method as
considered in Ref.~\cite{Gove71}. We used the natural units  $\hbar = m_{e} = c = 1$ in Eqs. (8) and (9). G$_{+}$ and G$_{-}$ represent the
positron and electron distribution functions, respectively, and are
expressed as
\begin{equation}
G_{+} =\left[\exp \left(\frac{E_{f} +E +2}{kT}\right)+1\right]^{-1},
\label{Gp}
\end{equation}
\begin{equation}
 G_{-} =\left[\exp \left(\frac{E - E_{f}}{kT}
 \right)+1\right]^{-1},
\label{Gm}
\end{equation}
where E = (w - 1) represents the electron K.E and $E_{f}$ denote the
electrons Fermi energy.

Due to the high stellar core temperature, there exist a definite
possibility of occupation of parent excited levels. The total
(anti)neutrino energy loss weak-rates per unit time per nuclide is
specified as
\begin{equation}
\lambda^{\nu(\bar{\nu})} =\sum _{nm}P_{n} \lambda
_{nm}^{\nu(\bar{\nu})}, \label{rate}
\end{equation}
where the probability of occupation of parent excited levels is
represented by $P_{n}$ , which obeys the normal Boltzmann
distribution. In Eq.~(\ref{rate}), the summation runs on all final
and initial levels until convergence is achieved in the rate
calculations.

\section{Results and Discussions} \label{sec:results}

Nickel (Ni) nuclide are considered to play a crucial role in the
evolutionary phases of massive stars. Previous simulation results
demonstrate that Ni nuclei ($^{56-61}$Ni, $^{63-69}$Ni and
$^{71}$Ni) appreciably alter the lepton fraction of the
astrophysical core of high mass stars \cite{Aufder94,Heger01a}. The
lepton capture and stellar $\beta$-weak rates are controlled by
Gamow-Teller (GT) charge-changing transitions (and to a much minor
extent by Fermi strength). The GT charge-changing transitions are of
spin-isospin ($\sigma\tau$) type and play significant role in
nuclear weak-decay routes. The charge-changing GT strength
distributions on $^{56-65}$Ni nuclei have been discussed in detail
recently (see \cite{sad18}). The authors in Ref.~\cite{sad18}, have
considered four different QRPA models and have compared their
calculated GT strength with previous theoretical work and
experimental data, where it was found that the current deformed
pn-QRPA model results are in decent accordance with experimental
data. For detailed description on GT transitions strength on
$^{56-65}$Ni, see Figs.~5-11 of Ref.~\cite{sad18}.
\begin{table}
  \centering
\caption{\small Comparison of pn-QRPA calculated $\beta$-decay
half-lives (T$_{1/2}$) with measured data \cite{Aud12} (in units of
second). The last two columns show the comparison of (re-normalized)
calculated and theoretical Ikeda sum rule of $^{56-71}$Ni,
respectively. $^{58}$Ni, $^{60-62}$Ni and $^{64}$Ni are stable
nuclide. }\label{Table ISR}
    \begin{tabular}{ccccc}
\hline
Nuclei & T$_{1/2}$ (Cal)& T$_{1/2}$ (Exp) & Re-ISR$_{Cal}$& Re-ISR$_{Th}$\\
\hline\\
$^{56}$Ni & 5.37$\times$ 10$^{5}$ & 5.25$\times$ 10$^{5}$ & 0.00  & 0.00 \\
 $^{57}$Ni & 1.31$\times$ 10$^{5}$ & 1.28$\times$ 10$^{5}$ & 1.08  & 1.08 \\
 $^{58}$Ni & - &      - & 2.16  & 2.16 \\
 $^{59}$Ni & 3.47$\times$ 10$^{12}$ & 3.19$\times$ 10$^{12}$ & 3.24  & 3.23 \\
 $^{60}$Ni & - &     -  & 4.32  & 4.32 \\
 $^{61}$Ni & - &    -   & 5.40  & 5.40 \\
 $^{62}$Ni & - &    -   & 6.47  & 6.47 \\
 $^{63}$Ni & 3.41$\times$ 10$^{9}$ & 3.19$\times$ 10$^{9}$ & 7.56  & 7.56 \\
 $^{64}$Ni & - &  -     & 8.64  & 8.64 \\
 $^{65}$Ni & 9.12$\times$ 10$^{3}$ & 9.06$\times$ 10$^{3}$ & 9.72  & 9.71 \\
 $^{66}$Ni & 2.11$\times$ 10$^{5}$ & 1.97$\times$ 10$^{5}$ & 10.80 & 10.80 \\
 $^{67}$Ni & 21.87 & 21.00 & 11.87 & 11.88 \\
 $^{68}$Ni & 30.61 & 29.00 & 12.96 & 12.96 \\
 $^{69}$Ni & 11.48 & 11.50 & 14.02 & 14.04 \\
 $^{70}$Ni & 6.04  & 6.00  & 15.12 & 15.12 \\
 $^{71}$Ni & 2.70  & 2.56  & 16.19 & 16.20 \\

\hline
\end{tabular}
\end{table}

The GT$_\pm$
transitions properties and the corresponding lepton capture rates,
on heavier $^{66-71}$Ni isotopes were recently calculated in
Ref.~\cite{Nab17a}. It has been observed that, experimentally
measured GT strength is generally less than that expected from
nuclear models. This observation applies to transitions between
single, defined energy levels as well as to entire GT strength
functions. This discrepancy in shell model calculations is normally
explained by high order configuration mixing and intrinsic
excitations of the nucleons. Most nuclear models then renormalize
the GT strength by some fixed quenching factor. For RPA calculation
a quenching factor ($f_{q}$) of 0.6 is mostly used (e.g. in
~\cite{Gaa83, Vet89, Roe93}) and is also employed in current
calculation.
The
relation for Ikeda sum rule (ISR$_{r-n}$), in re-normalized form, in
our model is specified as
\begin{equation}
ISR_{r-n} = \sum B(GT_{-}) - \sum B(GT_{+})\cong 3f_{q}^{2}(N-Z).
\label{Eqt. ISR}
\end{equation}

\begin{figure}
%\vspace{1cm}
\hspace{-0.4in}\includegraphics[height=1.2\textwidth,
width=1.4\textwidth]{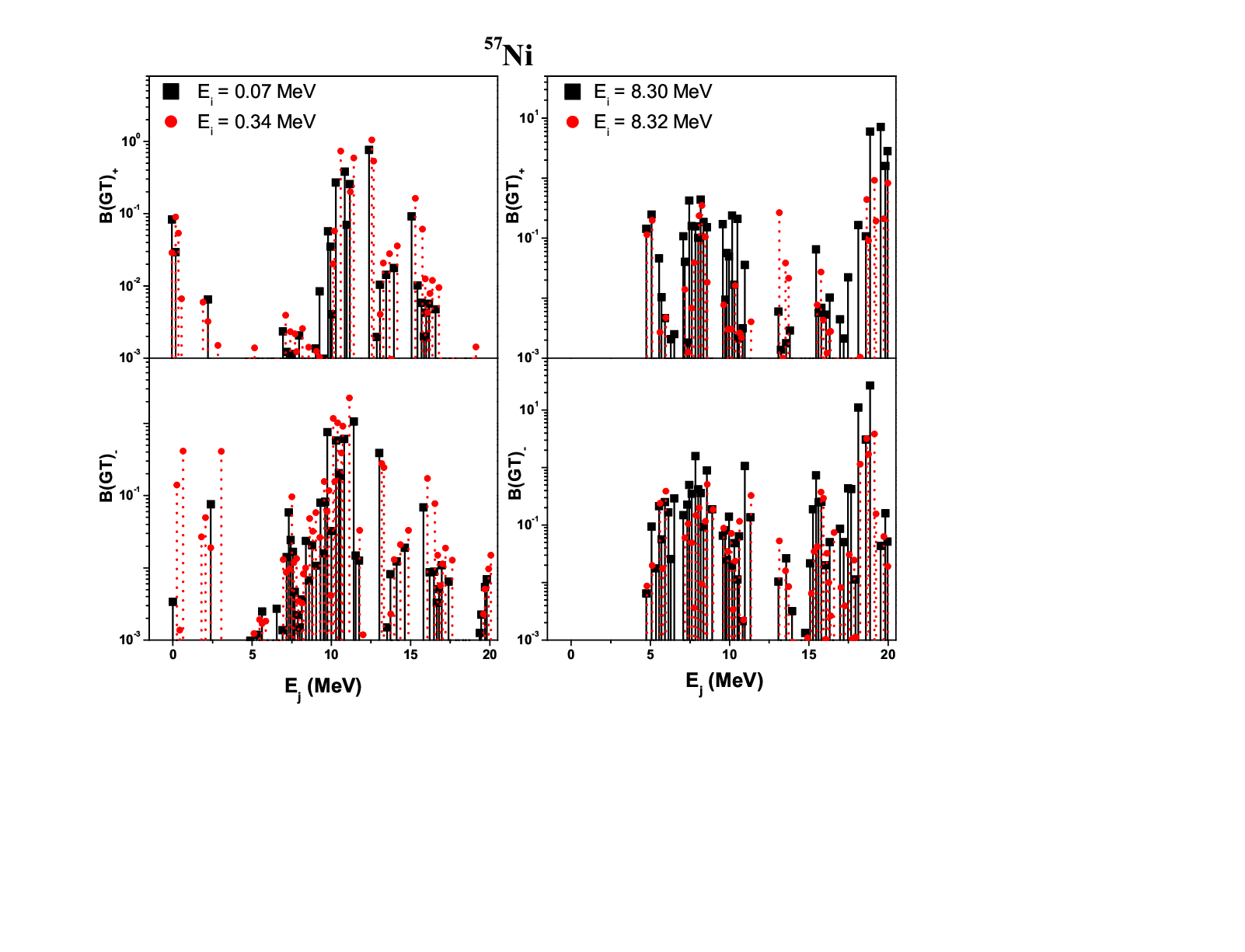} \vspace{-3cm}\caption{The pn-QRPA
calculated GT strength distributions as a function of daughter
excitation energy, $E_{j}$, along electron capture
($\beta^{+}$-decay)  (top panels) and positron capture
($\beta^{-}$-decay) directions (bottom panels) for $^{57}$Ni. The
left panels show GT transitions from low-lying excited states of the
parent nucleus, whereas right panels show GT transitions from
high-lying excited states of the parent nucleus.} \label{57-bgt}
\end{figure}

\begin{figure}
%\vspace{1cm}
\hspace{-0.4in}\includegraphics[height=1.2\textwidth,
width=1.4\textwidth]{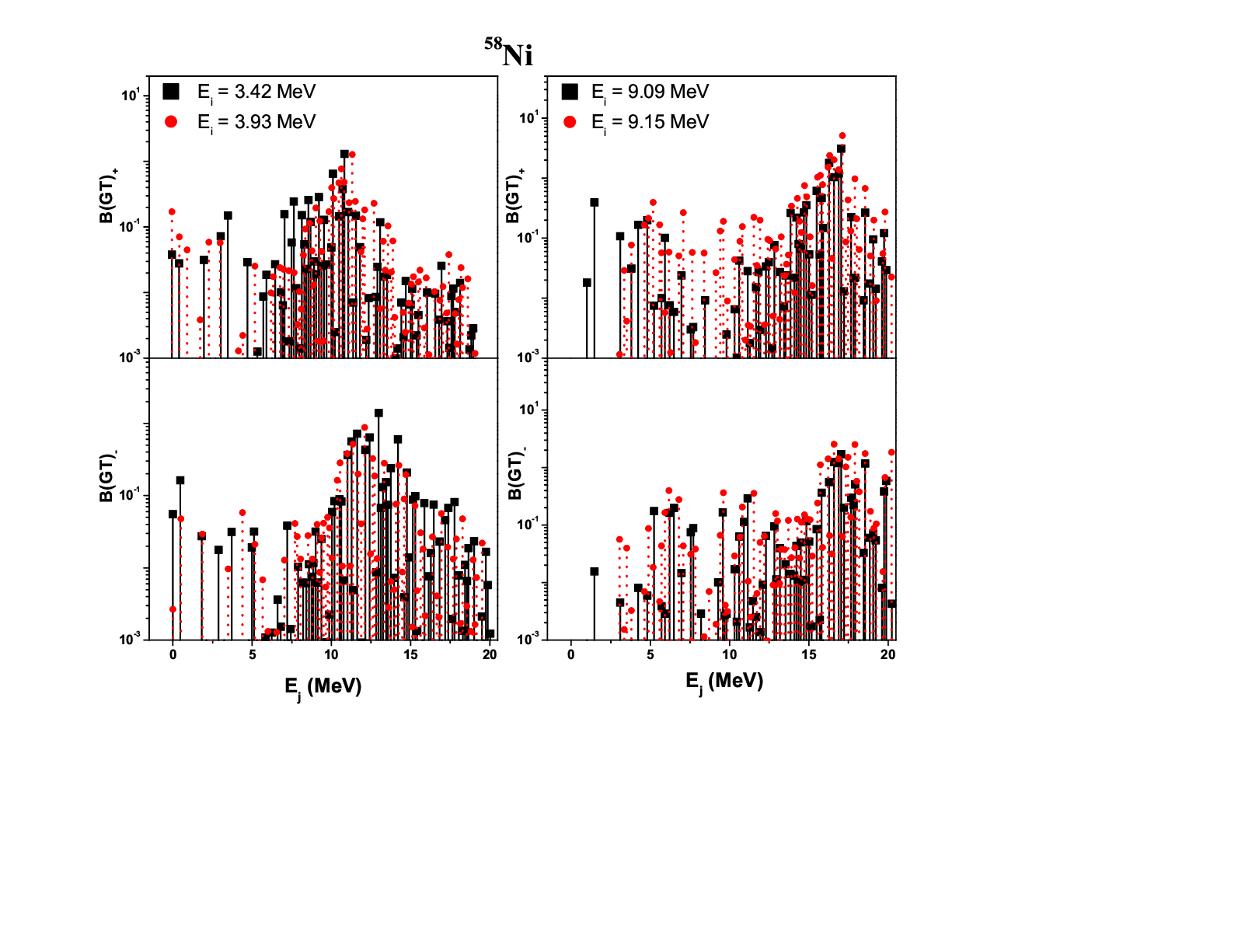} \vspace{-3cm}\caption{Same as
Figure~\ref{57-bgt}, but for $^{58}$Ni} \label{59-bgt}
\end{figure}

\begin{figure}
%\vspace{1cm}
\hspace{-0.4in}\includegraphics[height=1.2\textwidth,
width=1.4\textwidth]{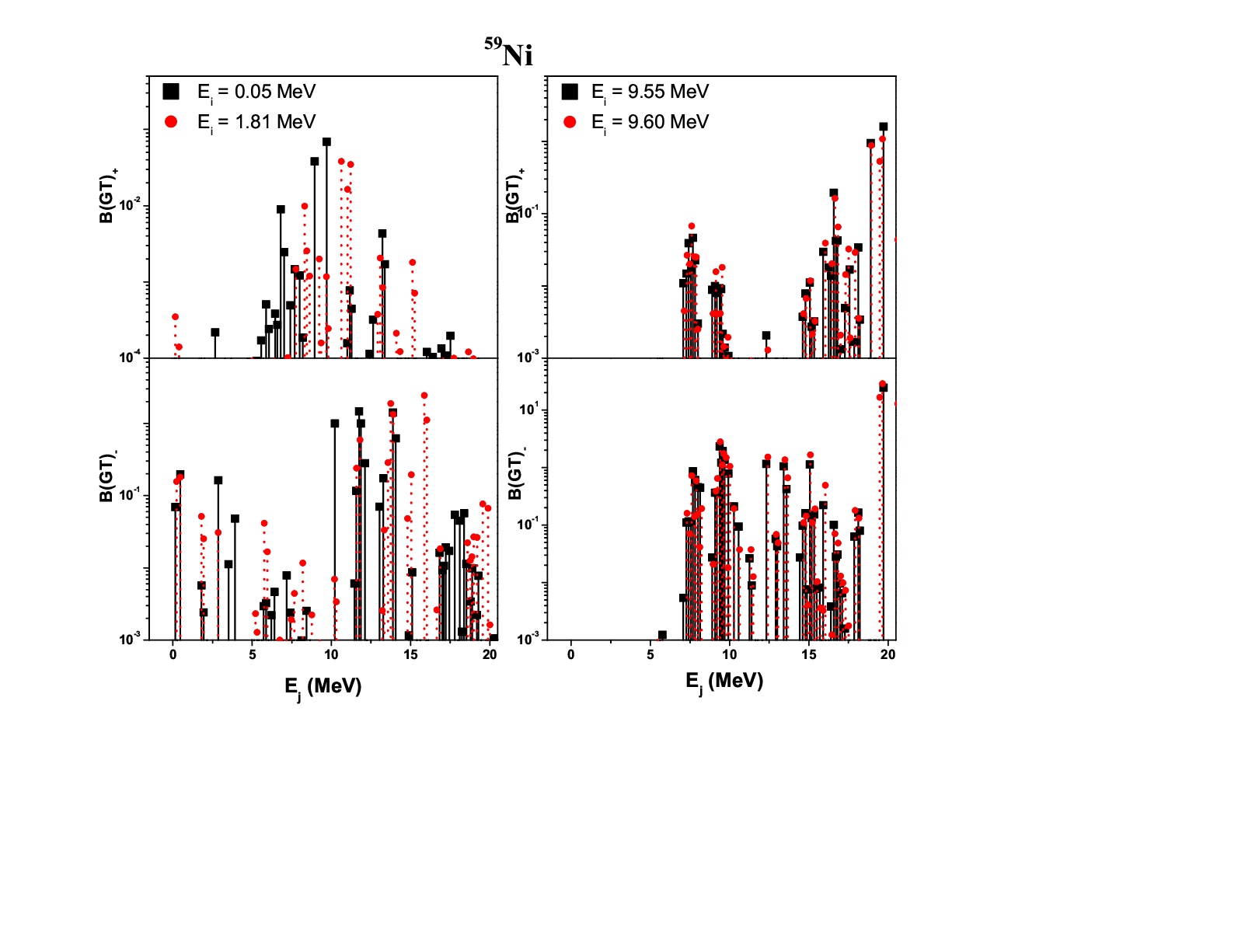} \vspace{-3cm}\caption{Same as
Figure~\ref{57-bgt}, but for $^{59}$Ni} \label{58-bgt}
\end{figure}

\begin{figure}
%\vspace{1cm}
\hspace{-0.4in}\includegraphics[height=1.2\textwidth,
width=1.4\textwidth]{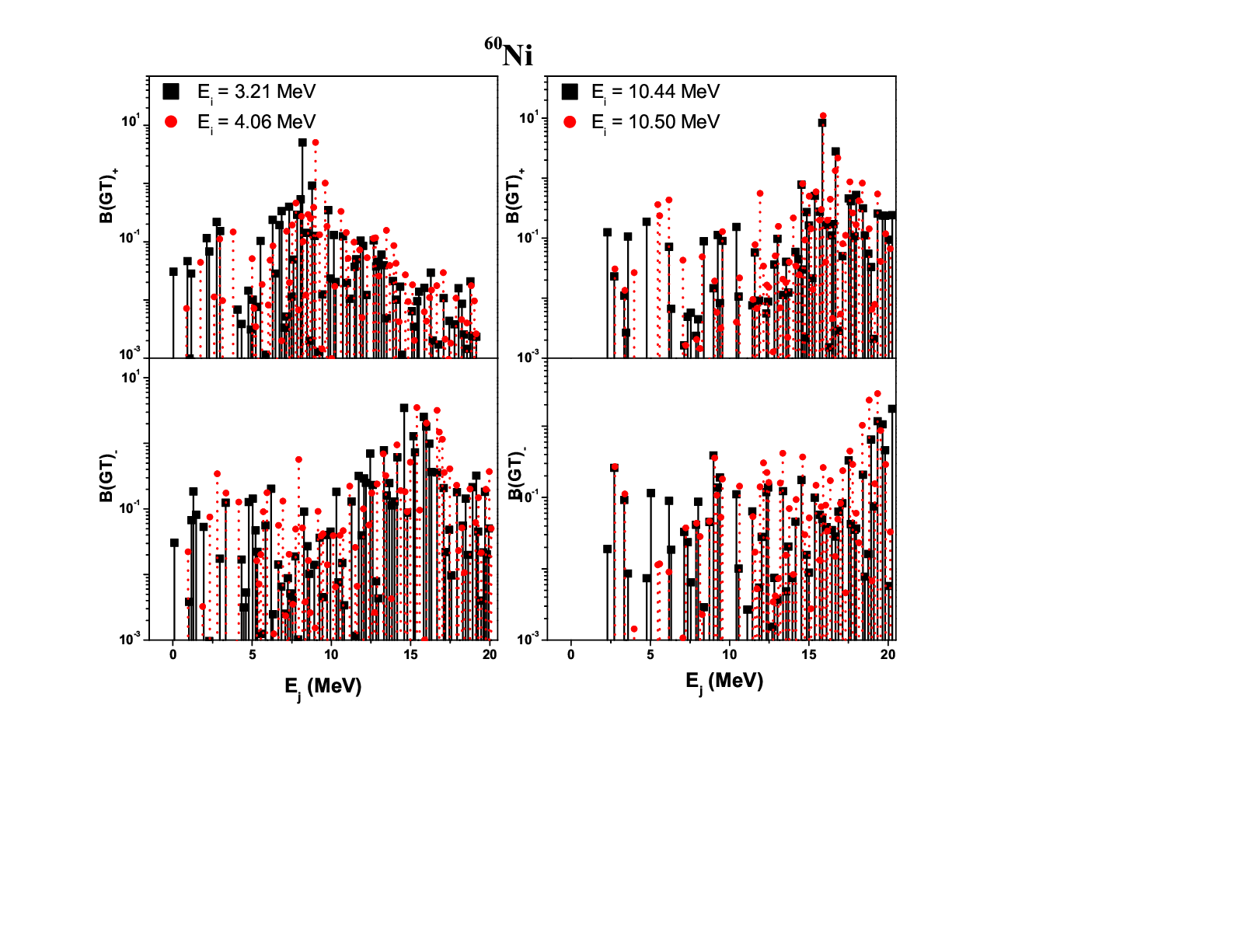} \vspace{-3cm}\caption{Same as
Figure~\ref{57-bgt}, but for $^{60}$Ni} \label{60-bgt}
\end{figure}

In this paper we would like to discuss the
calculated terrestrial half-lives (T$_{1/2}$) and the corresponding
(anti)neutrino cooling weak rates on Ni nuclide with mass ranging
from A = 56 to 71. These nuclide include both stable ($^{58}$Ni,
$^{60-62}$Ni and $^{64}$Ni) and unstable nuclei.
Table~\ref{Table ISR} depicts that our calculated T$_{1/2}$ values
are in excellent comparison with the measured data. The measured
T$_{1/2}$ values were chosen from \cite{Aud12}. As mentioned
earlier, $^{58}$Ni,
$^{60-62}$Ni and $^{64}$Ni are stable isotopes of nickel.\\
Table~\ref{Table ISR} demonstrate that our calculated ISR$_{r-n}$
values are in decent comparison with the theoretical predictions.
Thus the current calculations well satisfied the model independent ISR.\\
Our study suggests that the Brink-Axel hypothesis (BAH)  \cite{BAH}
is a poor estimation to be used in the computation of stellar
weak-decay rates. The pn-QRPA calculated GT strength distributions
for $^{57-60}$Ni, as a function of daughter excitation energy
($E_{j}$), along electron capture ($\beta^{+}$-decay) direction (top
panels) and positron capture ($\beta^{-}$-decay) direction (bottom
panels)  are shown in Figures~\ref{57-bgt}-\ref{60-bgt}. The left
panels show  GT strength distributions for low-lying excited states
of the parent nucleus, whereas right panels show GT strength
distributions for high-lying excited states of the parent nucleus.
It is to be noted that we calculate GT strength distributions from
all calculated parent states of all selected Ni isotopes and are not
shown here due to space consideration.  The ASCII files of all
allowed GT strength distributions may be requested from the
corresponding author. It can be seen that the calculated GT strength
distributions for excited states are different from each other. For
nickel isotopes these excited state GT strength distributions
contribute to the stellar weak-decay rates during the core
contraction and collapse phases of massive stars. The weak decay
rates are  exponentially sensitive to the position of Gamow-Teller
resonance \cite{Auf96}.

\begin{figure}
%\vspace{1cm}
\includegraphics[width=1\textwidth]{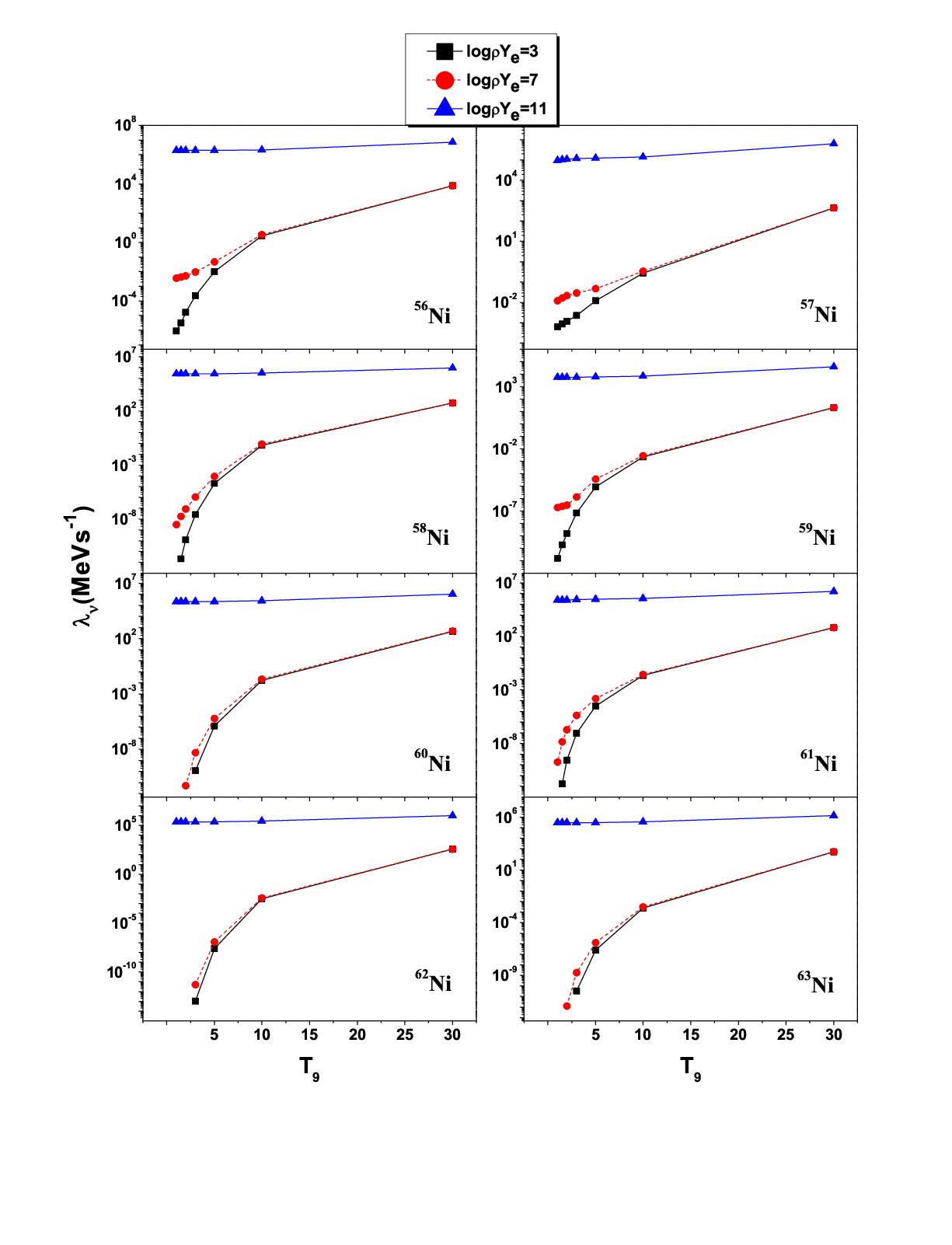}
\vspace{-3cm} \caption{The pn-QRPA calculated neutrino cooling rates
due to$~^{56-63}$Ni at different selected densities as a function of
stellar temperature. $\log\rho$Y$_{e}$ gives the $\log$ to base 10
of stellar density in units of g$\;$cm$^{-3}$. T$_{9}$ gives the
stellar temperature in units of $10^9\;$K and $\lambda_{\nu}$
represents neutrino cooling rates in units of MeV$\;$s$^{-1}$ (color
online).} \label{56-63-nu}
\end{figure}

\begin{center}
\begin{figure}
%\vspace{1cm}
\includegraphics[width=1\textwidth]{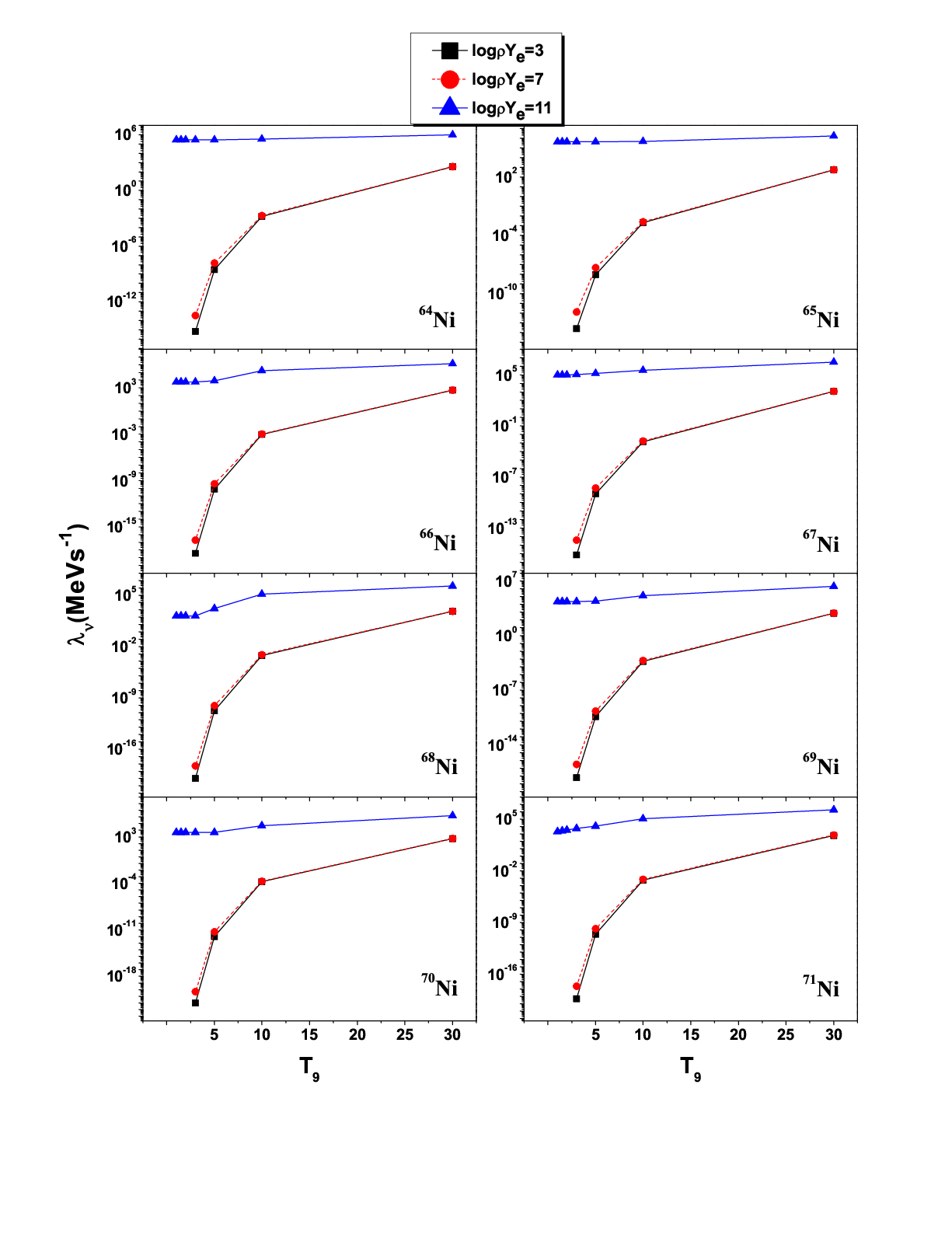}
\vspace{-3cm} \caption{Same as Fig.~\ref{56-63-nu} but for neutrino
cooling rates due to$~^{64-71}$Ni (color online).} \label{64-71-nu}
\end{figure}
\end{center}

\begin{center}
\begin{figure}
%\vspace{1cm}
\includegraphics[width=1\textwidth]{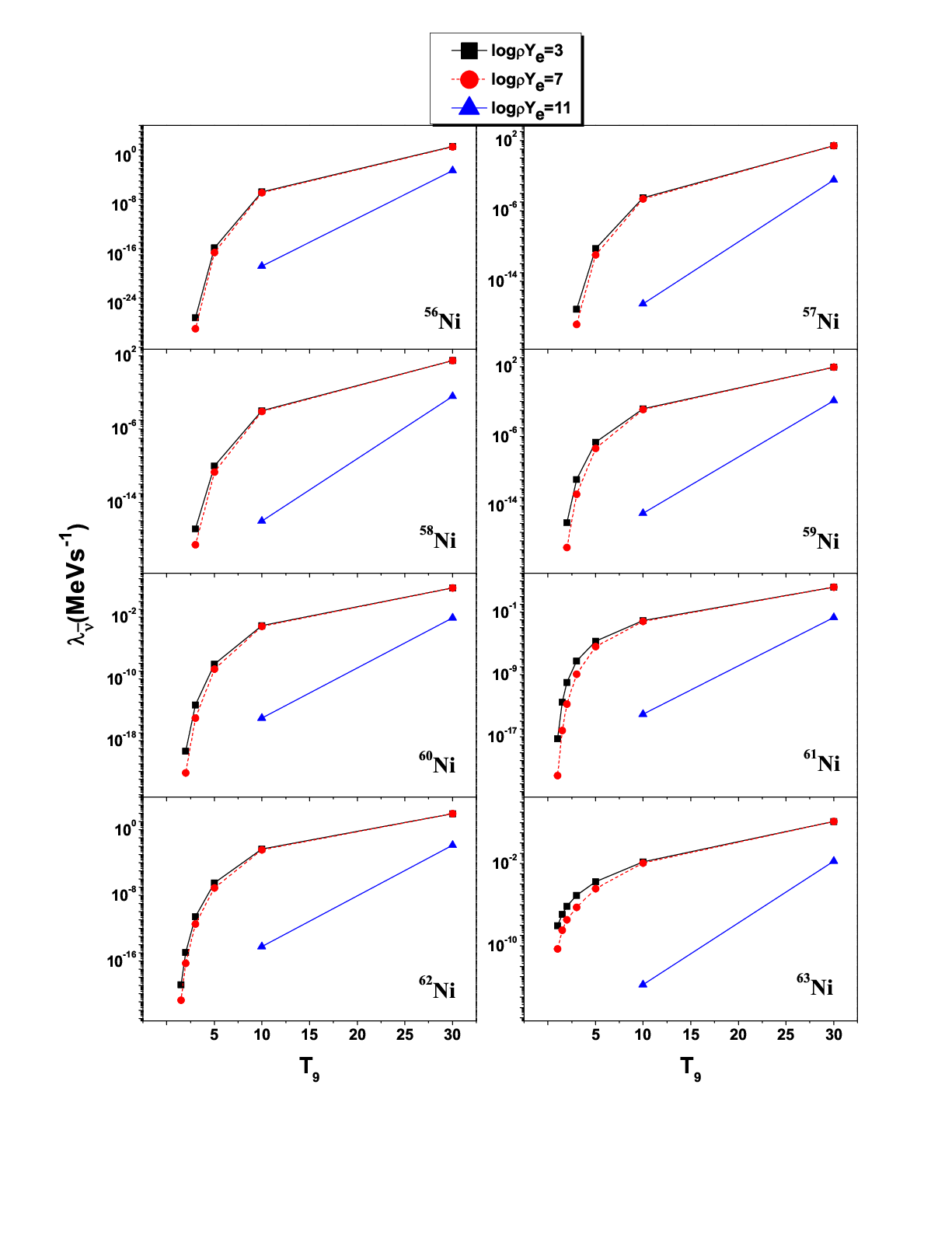}
\vspace{-3cm} \caption{Same as Fig.~\ref{56-63-nu} but for
antineutrino cooling rates due to$~^{56-63}$Ni (color online).}
\label{56-63-nubar}
\end{figure}
\end{center}

\begin{center}
\begin{figure}
%\vspace{1cm}
\includegraphics[width=1\textwidth]{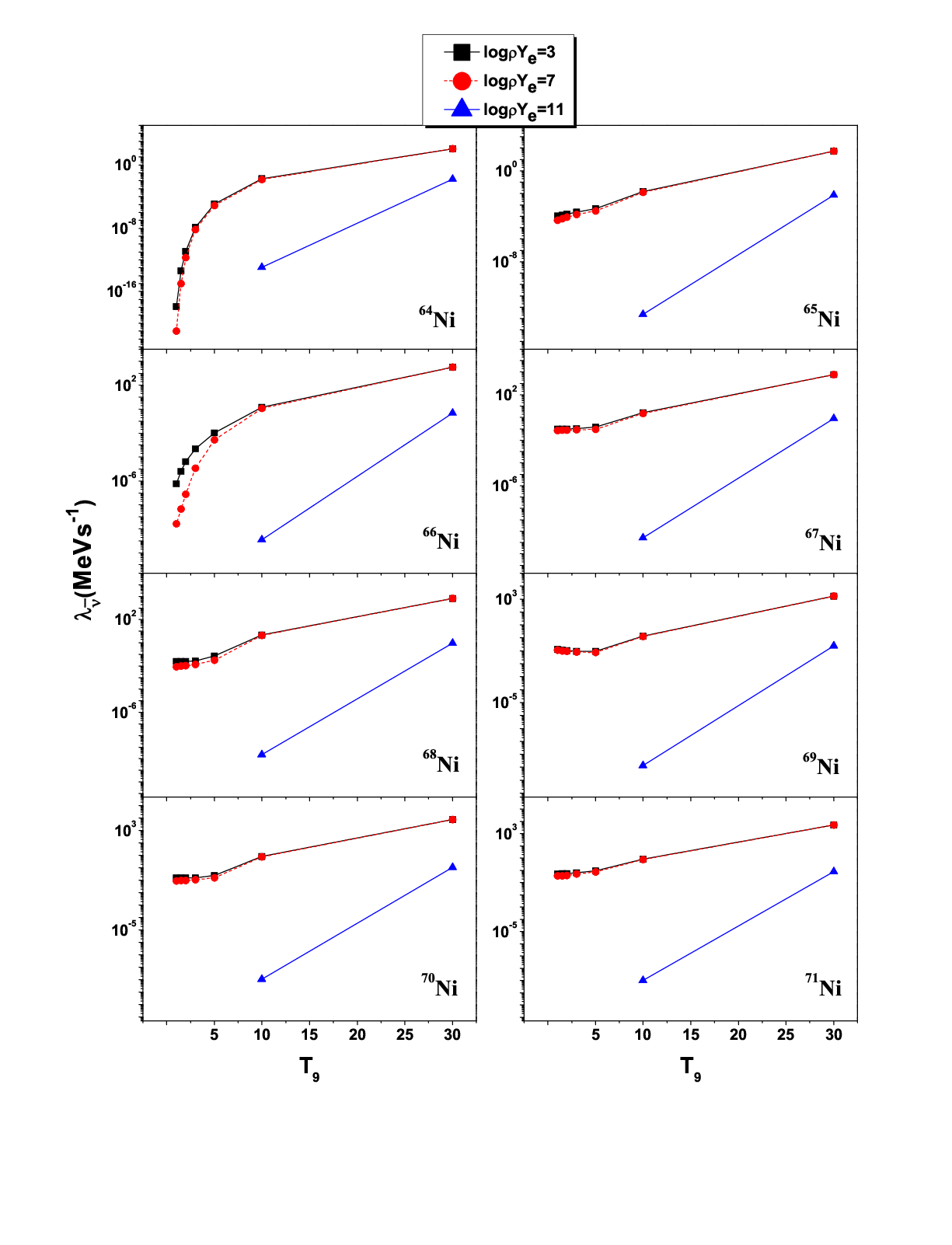}
\vspace{-3cm} \caption{Same as Fig.~\ref{56-63-nu} but for
antineutrino cooling rates due to$~^{64-71}$Ni (color online).}
\label{64-71-nubar}
\end{figure}
\end{center}

\begin{center}
\begin{figure}
%\vspace{1cm}
\includegraphics[width=0.9\textwidth]{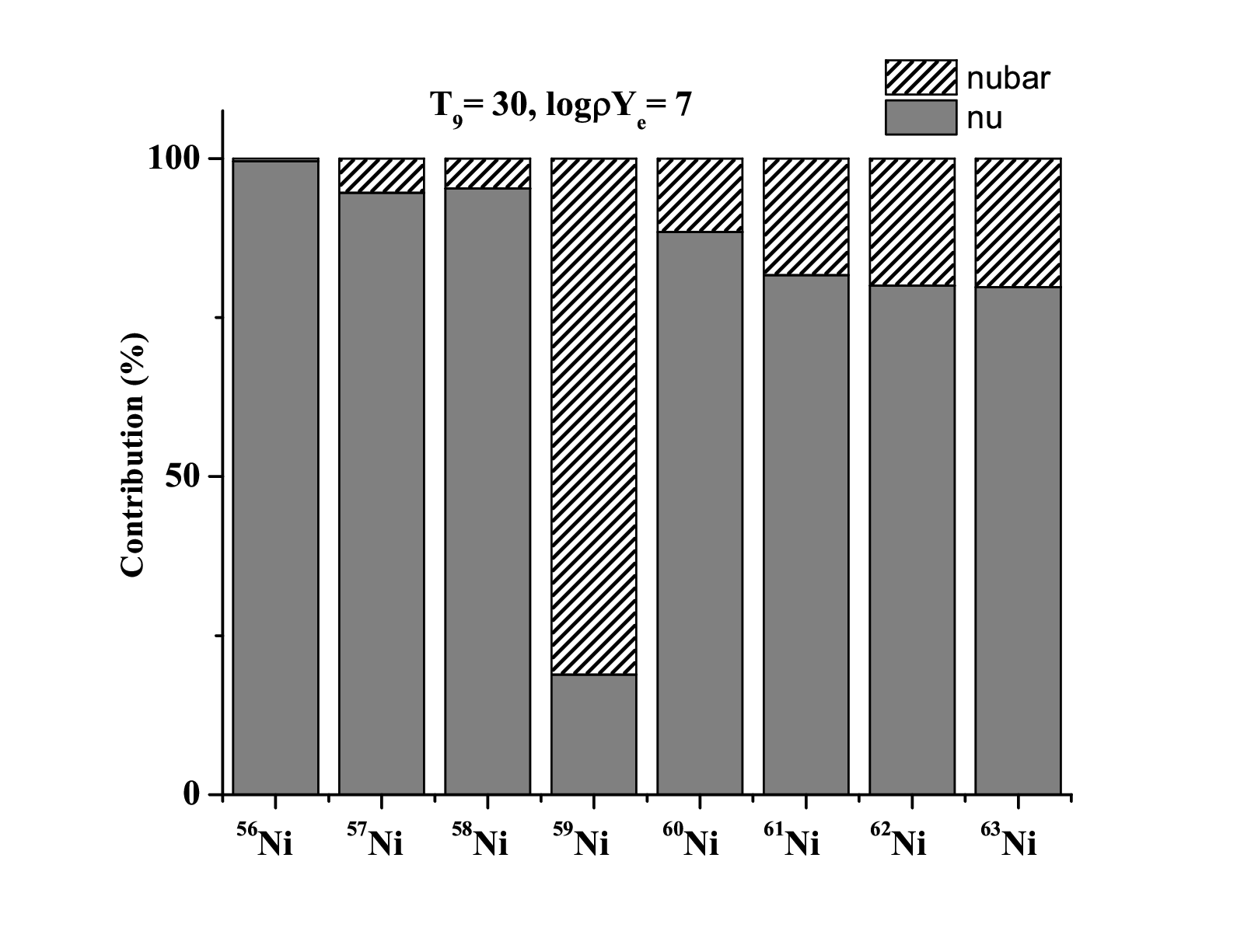}\\
\includegraphics[width=0.9\textwidth]{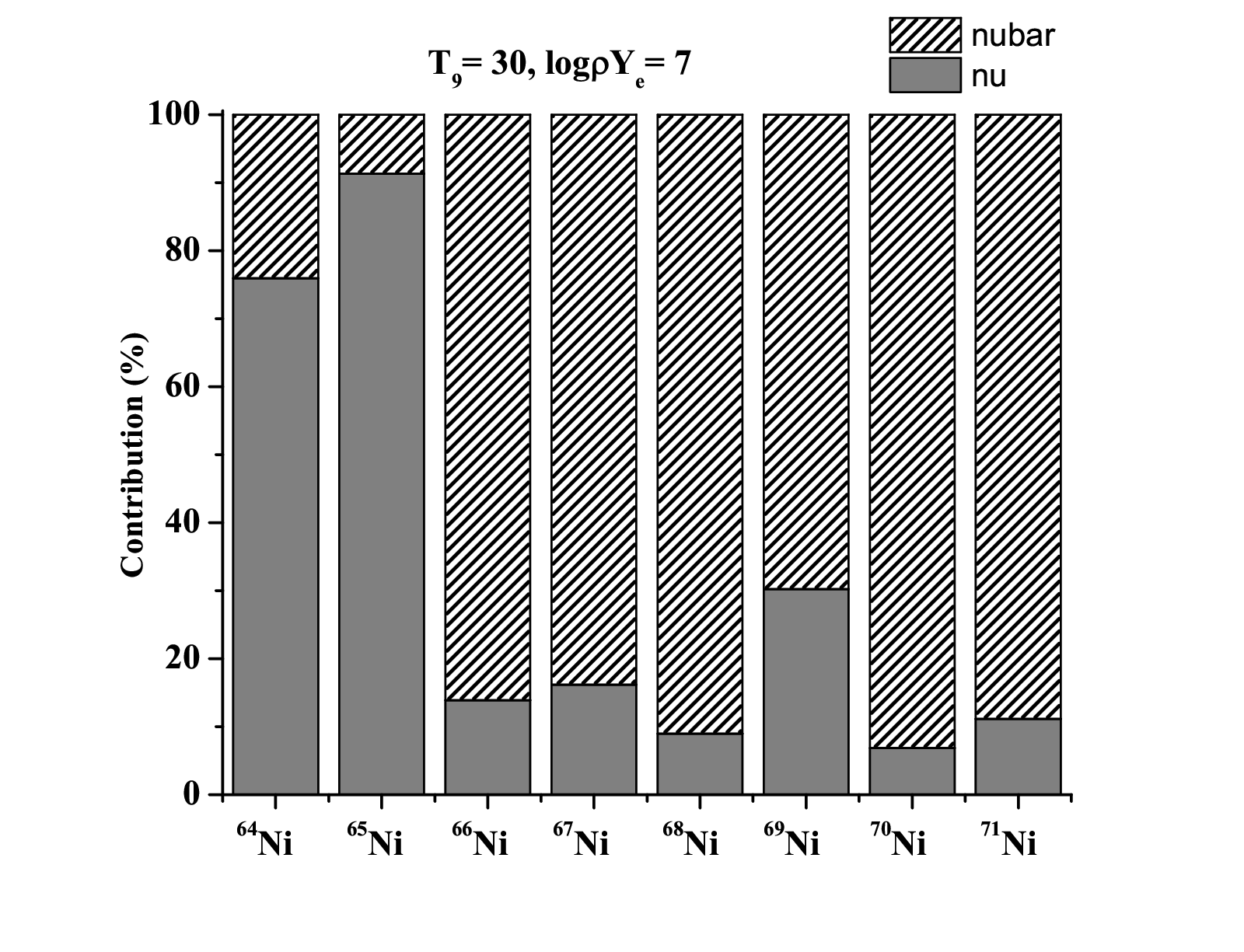}
\caption{Percentage contribution of neutrino and antineutrino cooling rates due to $~^{56-63}$Ni (top) and $~^{64-71}$Ni (bottom) at T$_{9} = 30$ and $\log\rho$Y$_{e} = 7$.} \label{nu-nubar-rho7-T30}
\end{figure}
\end{center}

The deformed pn-QRPA model (anti)neutrino cooling rates on
$^{56-71}$Ni isotopes are depicted in
Figures~\ref{56-63-nu}-\ref{64-71-nubar}. In each of these figures,
the calculated energy loss rates are presented as a function of
stellar temperature (T$_{9}$ in units of $10^{9}\;$K) ranging from 1
to 30, at three chosen values of stellar densities ($\log
\rho$Y$_{e}$ = 3, 7, 11, in units of g.cm$^{-3}$). These values
roughly represent low, medium and high stellar density regions.
$\lambda_{\nu}$ and $\lambda_{\bar{\nu}}$, represent total neutrino
and antineutrino cooling rates, respectively. $\lambda_{\nu}$
($\lambda_{\bar{\nu}}$) possesses contributions from positron
emissions and electron captures (positron captures and electron
emissions) on nickel isotopes. It can be seen from
Figures~\ref{56-63-nu}-\ref{64-71-nu}, with the rise of core
temperature up to T$_{9}=5$ in the low and medium density regions,
neutrino cooling rates increase exponentially. After this
temperature, there is a drastic reduction in the slope of the
weak-rates as the core density rises. The cooling rates due to
antineutrinos are shown in
Figures~\ref{56-63-nubar}-\ref{64-71-nubar}. With increasing core
temperature, there is also an increase in the antineutrino
weak-rates. However, the antineutrino weak-rates in the high density
region are orders of magnitude smaller than the ones in the domain
of low and medium densities, especially at low temperatures.
Figure~\ref{nu-nubar-rho7-T30} shows the relative contribution of
neutrino and antineutrino weak-rates for $^{56-71}$Ni nuclide at
$\log \rho$Y$_{e}$ = 7 and T$_{9}$ = 30.

Our calculated cooling rates on selected nickel isotopes are
also compared with earlier calculations done by LSSM \cite{Lang00},
FFN \cite{FFN} and PF \cite{Pruet03}.
Tables~\ref{table1}-\ref{table7} exhibit ratios of the calculated
values of (anti)neutrino weak-rates due to  $^{56-71}$Ni nuclide for
selected density and temperature values. In the first column of the
tables, the selected values of $\log \rho$Y$_{e}$ in units of
g/cm$^{3}$ are given. The second column specifies stellar
temperatures (T$_{9}$). R$_{\nu (\bar{\nu})}$(LSSM), R$_{\nu
(\bar{\nu})}$(FFN) and R$_{\nu (\bar{\nu})}$(PF) denote the ratios
of our computed rates to LSSM, FFN and PF rates,
respectively. For our list of nickel isotopes, for the mass range
A=56-60, comparison is made with both FFN and LSSM calculations. For
A=61-65 only LSSM rates, whereas from A=66 to 71, only PF rates were
available for the sake of comparison. As per studies of Aufderheide
et. al.~\cite{Aufder94}, $^{56}$Ni is amongst the top 3 $e^{-}$
capture nuclei that decrease Y$_{e}$. Table~\ref{table1} shows that,
for $^{56}$Ni, at low densities and low temperature (T$_{9}$ = 1),
the pn-QRPA estimated energy loss rates are larger as compared to
LSSM (FFN) computed rates by nearly a factor of 25 (24). As
temperature rises to T$_{9}$ = 30, and also at medium stellar
densities, our rates are still larger than the LSSM and FFN computed
rates by a factor of 2-6. At high stellar density, the mutual
comparison between the three model calculations improves to within a
factor 2. In case of $^{57}$Ni, at low temperatures, for low and
medium densities, our computed rates surpass LSSM and FFN computed
rates by up to 1-2 orders of magnitude, whereas at higher
temperatures (T$_{9} >$ 5) and in the high density region, LSSM
(FFN) results surpass our rates by a factor 3-11 (6-30). For $^{58,
59, 60}$Ni (see also Table~\ref{table2}), it is observed that in our selected domain of density
and temperature, LSSM and FFN calculated energy loss rates in
general exceed our rates. Excluding the case of $^{59}$Ni, the
agreement generally gets better at high temperature and in the high
density region. In case of $^{61}$Ni and $^{63}$Ni, the LSSM and
pn-QRPA neutrino loss rate calculations are in decent comparison
(within a factor of 3-4). For the cases of $^{62, 64, 65}$Ni (as in Table~\ref{table3}), at low
and medium densities and low temperatures, the LSSM rates are larger
than our calculated rates, whereas at high temperature and stellar
density, our computed rates get bigger.
%\begin{sidewaystable}[pt]
\begin{table}[pt]
\caption{\small Ratios of calculations of neutrino cooling rates due
to $^{56-58}$Ni at different selected densities and temperatures in
stellar matter. R$_{\nu}$(LSSM) and R$_{\nu}$(FFN) denote the ratios
of the pn-QRPA calculated neutrino cooling rates to those calculated
by large scale shell model (LSSM) \cite{Lang00} and those calculated
by Fuller and collaborators \cite{FFN} (FFN), respectively.
$\log\rho$Y$_{e}$ has units of g$\;$cm$^{-3}$, where $\rho$ is the
baryon density and Y$_{e}$ is the ratio of the lepton number to the
baryon number. Temperatures (T$_{9}$) are given in units of
$10^{9}\;$K.}\label{table1} \hspace{-1cm} {\small
\begin{tabular}{cccccccc}%{\textwidth}{c @{\extracolsep{\fill}} ccccccc}
& & & & & & & \\
\toprule
\multirow{2}{*}{$\log\rho$Y$_{e}$} &
\multirow{2}{*}{T$_{9}$} & \multicolumn{2}{c}{$^{56}$Ni}&
\multicolumn{2}{c}{$^{57}$Ni} & \multicolumn{2}{c}{$^{58}$Ni} \\
\cmidrule{3-4}  \cmidrule{5-6}  \cmidrule{7-8} & &
\multicolumn{1}{c}{R$_{\nu}$(LSSM)} &
\multicolumn{1}{c}{R$_{\nu}$(FFN)} &
\multicolumn{1}{c}{R$_{\nu}$(LSSM)} &
\multicolumn{1}{c}{R$_{\nu}$(FFN)} &
\multicolumn{1}{c}{R$_{\nu}$(LSSM)} &
\multicolumn{1}{c}{R$_{\nu}$(FFN)}  \\
\midrule
 3.0  &   1.00  &  2.48E+01  &  2.38E+01  &  4.56E+02  &  2.95E+02  &  2.75E+00  &  1.11E-01  \\
 3.0  &   1.50  &  6.31E+00  &  6.17E+00  &  1.67E+02  &  5.11E+01  &  3.46E-01  &  1.60E-02  \\
 3.0  &   2.00  &  4.80E+00  &  4.56E+00  &  5.62E+01  &  1.55E+01  &  1.66E-01  &  9.16E-03  \\
 3.0  &   3.00  &  3.97E+00  &  3.80E+00  &  1.65E+01  &  4.81E+00  &  1.17E-01  &  8.95E-03  \\
 3.0  &   5.00  &  2.47E+00  &  1.89E+00  &  3.60E+00  &  1.41E+00  &  1.48E-01  &  2.21E-02  \\
 3.0  &  10.00  &  2.36E+00  &  1.16E+00  &  4.04E-01  &  1.73E-01  &  3.71E-01  &  1.71E-01  \\
 3.0  &  30.00  &  5.75E+00  &  1.37E+00  &  3.85E-01  &  1.08E-01  &  7.60E-01  &  2.43E-01  \\
 & & & & & & & \\
 7.0  &   1.00  &  4.54E+00  &  4.31E+00  &  8.67E+00  &  8.24E+00  &  3.18E+01  &  6.00E+00  \\
 7.0  &   1.50  &  4.45E+00  &  4.29E+00  &  1.09E+01  &  9.51E+00  &  1.23E+00  &  1.09E-01  \\
 7.0  &   2.00  &  4.34E+00  &  4.28E+00  &  1.16E+01  &  8.00E+00  &  3.16E-01  &  2.60E-02  \\
 7.0  &   3.00  &  3.97E+00  &  3.86E+00  &  9.77E+00  &  4.31E+00  &  1.24E-01  &  1.14E-02  \\
 7.0  &   5.00  & 2.80E+00  &  1.95E+00  &  3.91E+00  &  1.46E+00  &  1.47E-01  &  2.07E-02  \\
 7.0  &  10.00  &  2.38E+00  &  1.16E+00  &  4.14E-01  &  1.75E-01  &  3.76E-01  &  1.71E-01  \\
 7.0  &  30.00  &  5.75E+00  &  1.37E+00  &  3.84E-01  &  1.08E-01  &  7.62E-01  &  2.43E-01  \\
  & & & & & & & \\
 11.0  &   1.00  &  1.63E+00  &  6.27E-01  &  8.95E-02  &  3.38E-02  &  4.45E-01  &  1.43E-01  \\
 11.0  &   1.50  &  1.63E+00  &  6.25E-01  &  9.59E-02  &  3.62E-02  &  4.46E-01  &  1.43E-01  \\
 11.0  &   2.00  &  1.62E+00  &  6.25E-01  &  1.02E-01  &  3.82E-02  &  4.45E-01  &  1.43E-01  \\
 11.0  &   3.00  &  1.62E+00  &  6.25E-01  &  1.11E-01  &  4.11E-02  &  4.44E-01  &  1.43E-01  \\
 11.0  &   5.00  & 1.62E+00  &  6.24E-01  &  1.22E-01  &  4.40E-02  &  4.39E-01  &  1.43E-01  \\
 11.0  &  10.00  &  1.75E+00  &  6.58E-01  &  1.35E-01  &  4.76E-02  &  4.59E-01  &  1.49E-01  \\
 11.0  &  30.00  &  4.12E+00  &  1.32E+00  &  3.92E-01  &  1.33E-01  &  8.71E-01  &  2.80E-01  \\
 \bottomrule
\end{tabular}}
\end{table}
%\end{sidewaystable}

\begin{table}[pt]
\caption{\small Same as Table~\ref{table1} but for $^{59-61}$Ni.}\label{table2}
\centering {\small
\begin{tabular}{cccccccc}%{\textwidth}{c @{\extracolsep{\fill}} ccccccc}
 & & & & & & \\
\toprule
\multirow{2}{*}{$\log\rho$Y$_{e}$} &
\multirow{2}{*}{T$_{9}$} & \multicolumn{2}{c}{$^{59}$Ni}&
\multicolumn{2}{c}{$^{60}$Ni} & $^{61}$Ni\\
\cmidrule{3-4}  \cmidrule{5-6}  \cmidrule{7-7}& &
\multicolumn{1}{c}{R$_{\nu}$(LSSM)} &
\multicolumn{1}{c}{R$_{\nu}$(FFN)} &
\multicolumn{1}{c}{R$_{\nu}$(LSSM)} &
\multicolumn{1}{c}{R$_{\nu}$(FFN)} &
\multicolumn{1}{c}{R$_{\nu}$(LSSM)}
\\ \midrule
 3.0  &   1.00  &  4.23E-03  &  1.87E-03  &  1.61E-01  &  1.53E-02  &  3.13E-01\\
 3.0  &   1.50  &  2.12E-03  &  8.13E-04  &  1.12E-01  &  1.13E-02  &  4.54E-01\\
 3.0  &   2.00  &  2.24E-03  &  7.01E-04  &  8.51E-02  &  9.48E-03  &  5.83E-01\\
 3.0  &   3.00  &  7.76E-03  &  1.57E-03  &  8.67E-02  &  1.17E-02  &  8.47E-01\\
 3.0  &   5.00  &  3.35E-02  &  5.16E-03  &  1.50E-01  &  2.76E-02  &  1.33E+00\\
 3.0  &  10.00  &  1.94E-02  &  7.73E-03  &  4.25E-01  &  1.29E-01  &  1.02E+00\\
 3.0  &  30.00  &  3.37E-02  &  1.08E-02  &  1.00E+00  &  2.83E-01  &  2.33E+00\\
 & & & & & &      \\
 7.0  &   1.00  &  7.05E-03  &  1.65E-03  &  2.68E-01  &  2.61E-02  &  3.57E-01\\
 7.0  &   1.50  &  2.60E-03  &  7.55E-04  &  1.24E-01  &  1.23E-02  &  4.72E-01\\
 7.0  &   2.00  &  1.82E-03  &  4.90E-04  &  7.82E-02  &  8.69E-03  &  5.87E-01\\
 7.0  &   3.00  &  4.26E-03  &  8.24E-04  &  7.38E-02  &  9.93E-03  &  8.22E-01\\
 7.0  &   5.00  &  3.06E-02  &  4.58E-03  &  1.48E-01  &  2.62E-02  &  1.32E+00\\
 7.0  &  10.00  &  1.95E-02  &  7.73E-03  &  4.28E-01  &  1.29E-01  &  1.03E+00\\
 7.0  &  30.00  &  3.37E-02  &  1.08E-02  &  1.01E+00  &  2.83E-01  &  2.34E+00\\
 & & & & & &       \\
 11.0  &   1.00  &  8.67E-03  &  2.96E-03  &  5.43E-01  &  1.29E-01  &  7.74E-01\\
 11.0  &   1.50  &  8.75E-03  &  2.96E-03  &  5.43E-01  &  1.29E-01  &  7.74E-01\\
 11.0  &   2.00  &  8.83E-03  &  2.95E-03  &  5.43E-01  &  1.29E-01  &  7.91E-01\\
 11.0  &   3.00  &  8.95E-03  &  2.95E-03  &  5.47E-01  &  1.29E-01  &  8.39E-01\\
 11.0  &   5.00  &  9.16E-03  &  2.98E-03  &  5.65E-01  &  1.29E-01  &  9.14E-01\\
 11.0  &  10.00  &  1.04E-02  &  3.38E-03  &  6.35E-01  &  1.44E-01  &  1.05E+00\\
 11.0  &  30.00  &  3.73E-02  &  1.22E-02  &  1.32E+00  &  3.37E-01  &  2.90E+00\\
 \bottomrule
\end{tabular}}
\end{table}

\begin{table}[pt]
\caption{\small Same as Table~\ref{table1} but for $^{62-65}$Ni.}\label{table3}
\centering {\small
\begin{tabular}{ccccccc}%{\textwidth}{c @{\extracolsep{\fill}} cccccccc}
 & & & & &  \\
\toprule  \multirow{2}{*}{$\log\rho$Y$_{e}$} &
\multirow{2}{*}{T$_{9}$} & $^{62}$Ni&
$^{63}$Ni & $^{64}$Ni & $^{65}$Ni\\
\cmidrule{3-3}  \cmidrule{4-4}  \cmidrule{5-5} \cmidrule{6-6} & &
\multicolumn{1}{c}{R$_{\nu}$(LSSM)} &
\multicolumn{1}{c}{R$_{\nu}$(LSSM)} &
\multicolumn{1}{c}{R$_{\nu}$(LSSM)} &
\multicolumn{1}{c}{R$_{\nu}$(LSSM)}
\\ \midrule
3.0 & 1.00 & 3.91E-02 & 3.48E-01 & 4.12E-02 & 4.19E-01\tabularnewline
3.0 & 1.50 & 4.55E-02 & 4.37E-01 & 6.92E-02 & 4.65E-01\tabularnewline
3.0 & 2.00 & 5.19E-02 & 5.00E-01 & 1.36E-01 & 5.07E-01\tabularnewline
3.0 & 3.00 & 6.64E-02 & 5.70E-01 & 3.27E-01 & 5.83E-01\tabularnewline
3.0 & 5.00 & 1.24E-01 & 5.32E-01 & 6.59E-01 & 9.57E-01\tabularnewline
3.0 & 10.00 & 4.59E-01 & 5.05E-01 & 1.51E+00 & 2.55E+00\tabularnewline
3.0 & 30.00 & 1.36E+00 & 2.61E+00 & 2.74E+00 & 6.44E+00\tabularnewline
 &  &  &  &  & \tabularnewline
7.0 & 1.00 & 2.03E-02 & 3.61E-01 & 4.14E-02 & 4.22E-01\tabularnewline
7.0 & 1.50 & 3.30E-02 & 4.43E-01 & 6.95E-02 & 4.69E-01\tabularnewline
7.0 & 2.00 & 4.44E-02 & 5.06E-01 & 1.36E-01 & 5.14E-01\tabularnewline
7.0 & 3.00 & 6.37E-02 & 5.77E-01 & 3.29E-01 & 5.87E-01\tabularnewline
7.0 & 5.00 & 1.26E-01 & 5.40E-01 & 6.65E-01 & 9.66E-01\tabularnewline
7.0 & 10.00 & 4.62E-01 & 5.07E-01 & 1.52E+00 & 2.56E+00\tabularnewline
7.0 & 30.00 & 1.37E+00 & 2.61E+00 & 2.74E+00 & 6.43E+00\tabularnewline
 &  &  &  &  & \tabularnewline
11.0 & 1.00 & 1.28E+00 & 1.67E+00 & 4.79E+00 & 5.27E+00\tabularnewline
11.0 & 1.50 & 1.28E+00 & 1.69E+00 & 4.80E+00 & 5.53E+00\tabularnewline
11.0 & 2.00 & 1.28E+00 & 1.71E+00 & 4.79E+00 & 5.66E+00\tabularnewline
11.0 & 3.00 & 1.28E+00 & 1.75E+00 & 4.75E+00 & 5.74E+00\tabularnewline
11.0 & 5.00 & 1.31E+00 & 1.80E+00 & 4.44E+00 & 5.64E+00\tabularnewline
11.0 & 10.00 & 1.45E+00 & 1.91E+00 & 4.53E+00 & 5.53E+00\tabularnewline
11.0 & 30.00 & 2.30E+00 & 3.90E+00 & 5.25E+00 & 9.38E+00\tabularnewline
 \bottomrule\end{tabular}}
\end{table}

\begin{table}[pt]
\caption{\small Same as Table~\ref{table1} but for $^{66-71}$Ni.
R$_{\nu}$(PF) shows the ratios of the pn-QRPA calculated neutrino
cooling rates to those calculated by Pruet and Fuller
(PF)~\cite{Pruet03}.}\label{table4} \hspace{-0.7cm} {\small
\begin{tabular}{cccccccc}%{\textwidth}{c @{\extracolsep{\fill}} cccccccc}
 & & & & & & & \\
\toprule  \multirow{2}{*}{$\log\rho$Y$_{e}$} &
\multirow{2}{*}{T$_{9}$} & $^{66}$Ni& $^{67}$Ni& $^{68}$Ni& $^{69}$Ni&
$^{70}$Ni & $^{71}$Ni \\
\cmidrule{3-3}  \cmidrule{4-4}  \cmidrule{5-5} \cmidrule{6-6} \cmidrule{7-7} \cmidrule{8-8}& &
\multicolumn{1}{c}{R$_{\nu}$(PF)} &
\multicolumn{1}{c}{R$_{\nu}$(PF)} &
\multicolumn{1}{c}{R$_{\nu}$(PF)} &
\multicolumn{1}{c}{R$_{\nu}$(PF)} &
\multicolumn{1}{c}{R$_{\nu}$(PF)} &
\multicolumn{1}{c}{R$_{\nu}$(PF)}
\\ \midrule
3.0 & 1.00 & 4.94E+00 & 5.09E+00 & 9.51E+02 & 8.30E+00 & 2.19E+02 & 1.73E+02\tabularnewline
3.0 & 1.50 & 7.73E+00 & 5.19E+00 & 9.57E+02 & 5.97E+00 & 2.96E+02 & 1.85E+02\tabularnewline
3.0 & 2.00 & 1.04E+01 & 6.18E+00 & 9.42E+02 & 4.93E+00 & 3.47E+02 & 1.90E+02\tabularnewline
3.0 & 3.00 & 1.60E+01 & 9.91E+00 & 8.67E+02 & 4.30E+00 & 3.91E+02 & 1.94E+02\tabularnewline
3.0 & 5.00 & 3.27E+01 & 1.86E+01 & 6.81E+02 & 6.32E+00 & 3.61E+02 & 1.87E+02\tabularnewline
3.0 & 10.00 & 2.28E+01 & 2.10E+01 & 2.54E+02 & 2.12E+01 & 8.95E+01 & 1.26E+02\tabularnewline
3.0 & 30.00 & 1.26E+01 & 1.44E+01 & 2.52E+01 & 1.10E+01 & 2.65E+01 & 1.53E+01\tabularnewline
 &  &  &  &  &  &  & \tabularnewline
7.0 & 1.00 & 4.98E+00 & 5.16E+00 & 9.59E+02 & 8.38E+00 & 2.21E+02 & 1.75E+02\tabularnewline
7.0 & 1.50 & 7.78E+00 & 5.21E+00 & 9.64E+02 & 6.01E+00 & 2.98E+02 & 1.86E+02\tabularnewline
7.0 & 2.00 & 1.04E+01 & 6.21E+00 & 9.46E+02 & 4.95E+00 & 3.48E+02 & 1.91E+02\tabularnewline
7.0 & 3.00 & 1.60E+01 & 9.91E+00 & 8.67E+02 & 4.30E+00 & 3.90E+02 & 1.94E+02\tabularnewline
7.0 & 5.00 & 3.30E+01 & 1.86E+01 & 6.84E+02 & 6.32E+00 & 3.68E+02 & 1.87E+02\tabularnewline
7.0 & 10.00 & 2.37E+01 & 2.13E+01 & 2.59E+02 & 2.14E+01 & 9.25E+01 & 1.27E+02\tabularnewline
7.0 & 30.00 & 1.26E+01 & 1.45E+01 & 2.52E+01 & 1.10E+01 & 2.64E+01 & 1.53E+01\tabularnewline
 &  &  &  &  &  &  & \tabularnewline
11.0 & 1.00 & 2.56E-01 & 2.85E+00 & 1.10E-02 & 7.18E-01 & 5.22E-01 & 1.23E-01\tabularnewline
11.0 & 1.50 & 2.56E-01 & 2.88E+00 & 1.10E-02 & 7.16E-01 & 5.24E-01 & 1.44E-01\tabularnewline
11.0 & 2.00 & 2.56E-01 & 2.87E+00 & 1.10E-02 & 7.24E-01 & 5.24E-01 & 1.86E-01\tabularnewline
11.0 & 3.00 & 2.62E-01 & 3.09E+00 & 1.16E-02 & 7.43E-01 & 5.42E-01 & 3.10E-01\tabularnewline
11.0 & 5.00 & 3.76E-01 & 4.12E+00 & 1.08E-01 & 7.96E-01 & 5.73E-01 & 6.08E-01\tabularnewline
11.0 & 10.00 & 6.50E+00 & 8.89E+00 & 9.27E+00 & 3.76E+00 & 4.63E+00 & 5.90E+00\tabularnewline
11.0 & 30.00 & 1.94E+01 & 2.82E+01 & 3.75E+01 & 2.06E+01 & 3.86E+01 & 2.63E+01\tabularnewline
 \bottomrule\end{tabular}}
\end{table}

The differences
seen in our and previous calculations (LSSM and FFN) are due to the
following reasons. Primarily, we did not consider the Brink-Axel
hypothesis (BAH) as used in shell model and FFN
work. The pn-QRPA model performs a state-by-state
calculation of astrophysical weak-rates from parent to daughter levels in a
microscopic way. Recent calculations \cite{Mis14,Joh15} have shown
that for a reliable estimation of stellar rates the BAH is a poor
approximation. In addition, the shell model calculation also
suffered with convergence problem (as pointed by Ref.
\cite{Pruet03}). On the other hand the FFN calculations had issues
with placement of GT transitions centroids. The pn-QRPA nuclear
model did not posses such types of problems and is fully microscopic
in nature.
\begin{table}[pt]
\caption{\small Same as Table~\ref{table1} but for antineutrino
cooling rates due to $^{57-60}$Ni.} \label{table5}
\hspace{-2cm}{\small
\begin{tabular}{ccccccccc}%{\textwidth}{c @{\extracolsep{\fill}} ccccccccc}
 & & & & & & & & \\
\toprule  \multirow{2}{*}{$\log\rho$Y$_{e}$} &
\multirow{2}{*}{T$_{9}$} & $^{57}$Ni&
\multicolumn{2}{c}{$^{58}$Ni} & \multicolumn{2}{c}{$^{59}$Ni} & \multicolumn{2}{c}{$^{60}$Ni}\\
\cmidrule{3-3}  \cmidrule{4-5}  \cmidrule{6-7} \cmidrule{8-9}& &
\multicolumn{1}{c}{R$_{\bar{\nu}}$(FFN)} &
\multicolumn{1}{c}{R$_{\bar{\nu}}$(LSSM)} &
\multicolumn{1}{c}{R$_{\bar{\nu}}$(FFN)} &
\multicolumn{1}{c}{R$_{\bar{\nu}}$(LSSM)} &
\multicolumn{1}{c}{R$_{\bar{\nu}}$(FFN)} &
\multicolumn{1}{c}{R$_{\bar{\nu}}$(LSSM)} &
\multicolumn{1}{c}{R$_{\bar{\nu}}$(FFN)}
\\ \midrule
3.0 & 1.00 & 3.83E-06 & 3.66E-04 & 1.74E-04 & 1.63E-03 & 1.25E-03 & 5.86E-03 & 1.95E-03\tabularnewline
3.0 & 1.50 & 3.51E-04 & 5.55E-03 & 2.89E-03 & 5.09E-02 & 3.80E-02 & 2.08E-02 & 5.89E-03\tabularnewline
3.0 & 2.00 & 2.93E-03 & 1.82E-02 & 9.95E-03 & 2.65E-01 & 1.82E-01 & 4.35E-02 & 1.14E-02\tabularnewline
3.0 & 3.00 & 1.95E-02 & 5.24E-02 & 2.90E-02 & 1.21E+00 & 6.58E-01 & 8.67E-02 & 2.12E-02\tabularnewline
3.0 & 5.00 & 6.40E-02 & 1.26E-01 & 6.19E-02 & 2.81E+00 & 1.09E+00 & 1.65E-01 & 4.23E-02\tabularnewline
3.0 & 10.00 & 1.16E-01 & 6.01E-01 & 1.98E-01 & 2.24E+00 & 7.03E-01 & 7.52E-01 & 1.86E-01\tabularnewline
3.0 & 30.00 & 2.36E-01 & 3.43E+00 & 2.19E-01 & 6.56E+00 & 6.49E-01 & 3.63E+00 & 3.51E-01\tabularnewline
 &  &  &  &  &  &  &  & \tabularnewline
7.0 & 1.00 & 3.82E-06 & 2.65E-04 & 1.74E-04 & 1.42E-03 & 1.25E-03 & 9.33E-03 & 3.94E-03\tabularnewline
7.0 & 1.50 & 3.52E-04 & 4.12E-03 & 2.88E-03 & 4.11E-02 & 3.81E-02 & 1.59E-02 & 7.03E-03\tabularnewline
7.0 & 2.00 & 2.91E-03 & 1.41E-02 & 9.68E-03 & 1.90E-01 & 1.82E-01 & 2.70E-02 & 1.25E-02\tabularnewline
7.0 & 3.00 & 1.86E-02 & 4.43E-02 & 2.54E-02 & 7.21E-01 & 6.55E-01 & 4.48E-02 & 2.22E-02\tabularnewline
7.0 & 5.00 & 6.03E-02 & 1.20E-01 & 5.48E-02 & 1.90E+00 & 1.06E+00 & 1.08E-01 & 4.17E-02\tabularnewline
7.0 & 10.00 & 1.15E-01 & 5.98E-01 & 1.96E-01 & 2.18E+00 & 6.70E-01 & 7.33E-01 & 1.76E-01\tabularnewline
7.0 & 30.00 & 2.36E-01 & 3.43E+00 & 2.19E-01 & 6.56E+00 & 6.50E-01 & 3.63E+00 & 3.51E-01\tabularnewline
 &  &  &  &  &  &  &  & \tabularnewline
11.0 & 1.00 & -- & -- & -- & -- & -- & -- & --\tabularnewline
11.0 & 1.50 & -- & -- & -- & 5.68E-03 & 7.28E-04 & -- & --\tabularnewline
11.0 & 2.00 & 6.25E-04 & 2.31E-03 & 9.71E-04 & 2.46E-03 & 3.13E-04 & 6.30E-04 & 5.48E-05\tabularnewline
11.0 & 3.00 & 5.51E-03 & 9.55E-03 & 4.39E-03 & 1.53E-02 & 1.87E-03 & 1.02E-03 & 7.89E-05\tabularnewline
11.0 & 5.00 & 3.00E-02 & 3.78E-02 & 1.99E-02 & 7.71E-02 & 8.61E-03 & 4.16E-03 & 2.73E-04\tabularnewline
11.0 & 10.00 & 1.00E-01 & 3.74E-01 & 1.64E-01 & 4.75E-01 & 4.34E-02 & 1.49E-01 & 8.85E-03\tabularnewline
11.0 & 30.00 & 2.34E-01 & 3.33E+00 & 2.17E-01 & 6.27E+00 & 6.07E-01 & 3.42E+00 & 3.27E-01\tabularnewline
\bottomrule\end{tabular}}
\end{table}

\begin{table}[pt] \caption{\small Same as
Table~\ref{table1} but for antineutrino cooling rates due to
$^{61-65}$Ni.}\label{table6} \centering {\small
\begin{tabular}{ccccccc}%{\textwidth}{c @{\extracolsep{\fill}} ccccccc}
 & & & & & &  \\
\toprule  \multirow{2}{*}{$\log\rho$Y$_{e}$} &
\multirow{2}{*}{T$_{9}$} & $^{61}$Ni&
{$^{62}$Ni} & {$^{63}$Ni} & {$^{64}$Ni} & {$^{65}$Ni}\\
\cmidrule{3-3}  \cmidrule{4-4}  \cmidrule{5-5} \cmidrule{6-6} \cmidrule{7-7}& &
\multicolumn{1}{c}{R$_{\bar{\nu}}$(LSSM)} &
\multicolumn{1}{c}{R$_{\bar{\nu}}$(LSSM)} &
\multicolumn{1}{c}{R$_{\bar{\nu}}$(LSSM)} &
\multicolumn{1}{c}{R$_{\bar{\nu}}$(LSSM)} &
\multicolumn{1}{c}{R$_{\bar{\nu}}$(LSSM)}
\\ \midrule
3.0 & 1.00 & 2.96E-02 & 3.44E-03 & 3.12E+00 & 5.97E-06 & 1.54E-01\tabularnewline
3.0 & 1.50 & 1.80E-01 & 9.42E-03 & 3.10E+00 & 2.14E-04 & 1.45E-01\tabularnewline
3.0 & 2.00 & 6.47E-01 & 2.07E-02 & 3.08E+00 & 1.50E-03 & 1.58E-01\tabularnewline
3.0 & 3.00 & 1.96E+00 & 5.22E-02 & 3.20E+00 & 1.79E-02 & 1.89E-01\tabularnewline
3.0 & 5.00 & 2.96E+00 & 1.48E-01 & 1.82E+00 & 1.07E-01 & 1.49E-01\tabularnewline
3.0 & 10.00 & 1.96E+00 & 8.13E-01 & 6.89E-01 & 5.06E-01 & 1.08E-01\tabularnewline
3.0 & 30.00 & 6.85E+00 & 3.85E+00 & 5.02E+00 & 2.66E+00 & 5.93E-01\tabularnewline
 &  &  &  &  &  & \tabularnewline
7.0 & 1.00 & 2.85E-01 & 6.03E-01 & 9.75E-01 & 1.95E-03 & 1.34E-01\tabularnewline
7.0 & 1.50 & 3.13E-01 & 6.03E-01 & 1.01E+00 & 2.49E-02 & 1.33E-01\tabularnewline
7.0 & 2.00 & 6.30E-01 & 4.53E-01 & 1.15E+00 & 1.06E-01 & 1.52E-01\tabularnewline
7.0 & 3.00 & 1.45E+00 & 2.02E-01 & 1.23E+00 & 2.38E-01 & 1.97E-01\tabularnewline
7.0 & 5.00 & 1.75E+00 & 1.18E-01 & 8.41E-01 & 1.69E-01 & 1.69E-01\tabularnewline
7.0 & 10.00 & 1.82E+00 & 7.59E-01 & 6.07E-01 & 4.69E-01 & 1.04E-01\tabularnewline
7.0 & 30.00 & 6.84E+00 & 3.85E+00 & 5.01E+00 & 2.66E+00 & 5.93E-01\tabularnewline
 &  &  &  &  &  & \tabularnewline
11.0 & 1.00 & -- & -- & -- & -- & --\tabularnewline
11.0 & 1.50 & 3.45E-04 & 6.65E-03 & 2.47E-03 & 1.93E-02 & 1.49E-02\tabularnewline
11.0 & 2.00 & 8.30E-04 & 6.00E-03 & 2.58E-03 & 1.62E-02 & 1.62E-02\tabularnewline
11.0 & 3.00 & 3.28E-03 & 5.25E-03 & 2.51E-03 & 1.33E-02 & 1.77E-02\tabularnewline
11.0 & 5.00 & 1.31E-02 & 6.53E-03 & 2.49E-03 & 1.62E-02 & 1.77E-02\tabularnewline
11.0 & 10.00 & 1.14E-01 & 6.37E-02 & 1.31E-02 & 6.71E-02 & 2.07E-02\tabularnewline
11.0 & 30.00 & 6.17E+00 & 3.48E+00 & 3.98E+00 & 2.22E+00 & 5.37E-01\tabularnewline
\bottomrule\end{tabular}}
\end{table}

\begin{table}[pt]
\caption{\small Same as Table~\ref{table4} but for antineutrino
cooling rates due to $^{66-71}$Ni.}\label{table7} \hspace{-1cm}
{\small
\begin{tabular}{cccccccc}%{\textwidth}{c @{\extracolsep{\fill}} cccccccc}
 & & & & & & & \\
\toprule  \multirow{2}{*}{$\log\rho$Y$_{e}$} &
\multirow{2}{*}{T$_{9}$} & $^{66}$Ni&
{$^{67}$Ni} & {$^{68}$Ni} & {$^{69}$Ni} & {$^{70}$Ni} & {$^{71}$Ni}\\
\cmidrule{3-3}  \cmidrule{4-4}  \cmidrule{5-5} \cmidrule{6-6} \cmidrule{7-7} \cmidrule{8-8} & &
\multicolumn{1}{c}{R$_{\bar{\nu}}$(PF)} &
\multicolumn{1}{c}{R$_{\bar{\nu}}$(PF)} &
\multicolumn{1}{c}{R$_{\bar{\nu}}$(PF)} &
\multicolumn{1}{c}{R$_{\bar{\nu}}$(PF)} &
\multicolumn{1}{c}{R$_{\bar{\nu}}$(PF)} &
\multicolumn{1}{c}{R$_{\bar{\nu}}$(PF)}
\\ \midrule
3.0 & 1.00 & 9.91E-01 & 1.33E+00 & 8.18E-01 & 1.16E+00 & 7.26E-01 & 4.53E-01\tabularnewline
3.0 & 1.50 & 3.11E+00 & 1.15E+00 & 8.20E-01 & 8.38E-01 & 7.24E-01 & 3.71E-01\tabularnewline
3.0 & 2.00 & 3.79E+00 & 9.82E-01 & 8.13E-01 & 6.27E-01 & 7.14E-01 & 3.05E-01\tabularnewline
3.0 & 3.00 & 3.97E+00 & 6.90E-01 & 6.22E-01 & 3.69E-01 & 4.35E-01 & 2.14E-01\tabularnewline
3.0 & 5.00 & 2.79E+00 & 3.71E-01 & 1.63E-01 & 1.39E-01 & 5.85E-02 & 1.22E-01\tabularnewline
3.0 & 10.00 & 2.09E+00 & 9.38E-01 & 4.84E-01 & 1.50E-01 & 2.21E-01 & 2.90E-01\tabularnewline
3.0 & 30.00 & 1.41E+01 & 1.57E+01 & 1.75E+01 & 3.09E+00 & 1.41E+01 & 7.80E+00\tabularnewline
 &  &  &  &  &  &  & \tabularnewline
7.0 & 1.00 & 5.96E-01 & 1.28E+00 & 6.37E-01 & 1.47E+00 & 5.36E-01 & 3.16E-01\tabularnewline
7.0 & 1.50 & 2.34E-01 & 1.10E+00 & 6.59E-01 & 9.91E-01 & 5.48E-01 & 2.66E-01\tabularnewline
7.0 & 2.00 & 2.58E-01 & 9.33E-01 & 6.71E-01 & 7.00E-01 & 5.52E-01 & 2.29E-01\tabularnewline
7.0 & 3.00 & 8.99E-01 & 6.38E-01 & 4.56E-01 & 3.82E-01 & 3.34E-01 & 1.73E-01\tabularnewline
7.0 & 5.00 & 1.21E+00 & 2.77E-01 & 8.26E-02 & 1.28E-01 & 3.96E-02 & 1.00E-01\tabularnewline
7.0 & 10.00 & 1.89E+00 & 8.43E-01 & 4.72E-01 & 1.41E-01 & 2.11E-01 & 2.64E-01\tabularnewline
7.0 & 30.00 & 1.41E+01 & 1.57E+01 & 1.75E+01 & 3.10E+00 & 1.41E+01 & 7.80E+00\tabularnewline
 &  &  &  &  &  &  & \tabularnewline
11.0 & 1.00 & -- & -- & -- & 2.01E+04 & -- & 7.13E+10\tabularnewline
11.0 & 1.50 & 5.30E+05 & 2.30E+14 & 1.62E+11 & 1.36E+20 & 3.84E+14 & 2.69E+24\tabularnewline
11.0 & 2.00 & 7.29E+04 & 4.12E+10 & 6.31E+08 & 7.35E+14 & 1.67E+11 & 8.75E+17\tabularnewline
11.0 & 3.00 & 7.93E+03 & 5.79E+06 & 1.95E+06 & 3.16E+09 & 6.92E+07 & 2.03E+11\tabularnewline
11.0 & 5.00 & 9.33E-01 & 1.35E-01 & 1.73E+01 & 1.42E-01 & 5.33E-01 & 8.04E-03\tabularnewline
11.0 & 10.00 & 6.19E-01 & 2.79E-01 & 2.24E+00 & 1.09E-01 & 8.07E-01 & 7.78E-02\tabularnewline
11.0 & 30.00 & 1.39E+01 & 1.52E+01 & 1.76E+01 & 2.82E+00 & 1.43E+01 & 5.86E+00\tabularnewline
\bottomrule\end{tabular}}
\end{table}

Table~\ref{table4} shows the comparison of our estimated neutrino
cooling rates on $^{66-71}$Ni nuclide with the ones calculated by
PF. For $^{66, 69}$Ni, at low and medium density, our calculated
rates are bigger than PF rates. At high stellar density, the two
rates are in reasonable comparison. In case of $^{67}$Ni, our
reported rates are factor 3-28 bigger than PF rates for the entire
domain of selected density and temperature. For $^{68, 70, 71}$Ni,
the pn-QRPA and PF comparison shows similar trend, where in the low
and medium density regions pn-QRPA computed energy loss rates are
greater by up to 2 orders of magnitude than PF rates. At high
stellar temperature (where the occupation probability of parent
excited states increases) our computed rates are enhanced by an
order of magnitude as compared to PF rates. For these neutron rich
nuclide our rates are enhanced due to the reason that we have used a
large model space in pn-QRPA calculations (up to seven major
oscillator shells) which can effectively handle all excited levels
both in parent and daughter nuclide.

Now we move towards the comparison of antineutrino energy loss rates due
to $^{57-71}$Ni isotopes with the previous calculations, which has
been presented in Tables~$\ref{table5}-\ref{table7}$. The
antineutrino cooling rates in the high density domain are orders of
magnitude smaller than the ones in the low and medium density
regions, especially at low temperatures. Generally, smaller the
results for energy loss rates, larger the differences are observed
with the previous calculations. At $\log \rho$Y$_{e}$ = 11, for
temperatures, T$_{9} \leq 1$, the rates are reported to have values
even lower than $10^{-100}\;$MeV$\;s^{-1}$. The comparison is not
made for such low values of the rates. With the increasing stellar
temperature, the magnitude of the estimated energy loss rates
increases and the comparison between the different calculations
improves. For $^{57-60}$Ni, the FFN rates are available for the
comparison. At low temperatures, FFN rates exceed pn-QRPA computed
rates by up to 2-5 orders of magnitude for different isotopes. The
comparison between the results of two models improves with the
increasing temperature, but the FFN rates are still factor 2-4
larger than our rates at $T_{9} = 30$. For $^{58-65}$Ni, our
estimated energy loss rates are also compared with the ones
calculated by LSSM. In case of even isotopes ($^{58, 60, 62,
64}$Ni), for temperatures, $1 \leq $T$_{9} \leq 10$, LSSM computed
rates surpass our rates, whereas at T$_{9} = 30$, our estimated
rates are factor $\sim$3-4 larger. The comparison of pn-QRPA and
LSSM computed antineutrino cooling rates due to $^{59, 61, 63}$Ni is
fairly good for temperatures, $3 \leq $T$_{9} \leq 10$ in the domain
of low and medium densities. In the high density region, LSSM
calculated rates exceed by up to 3 orders of magnitude at lower
temperatures. At T$_{9} = 30$, our rates get bigger by up to a
factor of 7. In case of $^{65}$Ni, LSSM estimated cooling rates
exceed our rates in all regions of selected density and temperature.
At T$_{9} = 30$, the agreement between the two results is better but
still LSSM rates are factor two bigger. These differences in the
comparison appears due to the reasons discussed earlier.

For the next six isotopes, $^{66-71}$Ni, the pn-QRPA estimated cooling rates
have been compared with PF calculations. As can be noted from
Table$~\ref{table7}$, for these neutron rich isotopes, in the low and medium
density domain and at low temperatures, the comparison of our calculated
cooling rates is fairly good against the ones made by PF. At high stellar
density, where the calculated cooling rates are very small, large differences
(up to several orders of magnitude) in the two results can be observed
especially at low temperatures (T$_{9} \leq 3$). At higher temperatures, the
mutual comparison between the two results improves. At  T$_{9}$ = 30,
our rates are larger by up to an order of magnitude as in case of neutrino
rates.

\section{Conclusions} \label{sec:conclusions}
Gamow-Teller (GT) charge-changing transitions on nickel nuclide play a
crucial role in the evolutionary stages of massive stars. Lepton capture and
decay rates on nickel nuclide notably affect the lepton-to-baryon fraction.
Neutrinos and antineutrinos, produced in weak-decay reactions, are
transparent to stellar core and escape from the stellar interior.
(Anti)neutrinos reduce the stars entropy and take away energy, thereby cools
the stellar interior. The deformed pn-QRPA theory having a decent track
record of terrestrial half-lives values was employed to predict the
(anti)neutrinos cooling rates on nickel nuclide in stellar scenario. Our
calculated beta decay half-lives for selected nickel isotopes are in
excellent comparison with measured values.

We have computed the neutrino and antineutrino cooling rates over
broad range of densities (10 -- 10$^{11}$g/cm$^{3}$) and stellar
temperatures (0.01$\times$10$^{9}$ -- 30$\times$10$^{9}$K) domains.
The comparison of computed neutrino cooling rates with the earlier
theoretical work of FFN and LSSM (wherever possible) was done. It
should be noted that the calculations of FFN and LSSM employed the
Brink-Axel hypothesis, not adopted in our pn-QRPA calculation. The
LSSM calculation also suffered with convergence problem (as pointed
by Ref. \cite{Pruet03}). On the other hand FFN calculations had
issues with placement of GT transitions centroids. Our nuclear model
did not encounter these problems. The current calculations are fully
microscopic in nature. For neutron-rich nickel nuclide we compared
our results with that of Pruet and Fuller work. At  high stellar
temperature (T$_{9}$ = 30) our rates are higher by up to an order of
magnitude than PF. Specially for neutron rich nickel isotopes, at
high core temperatures, our computed cooling rates are enhanced as
compare to previous calculations. We recommend core-collapse
simulators to test run our computed neutrino cooling rates which we
believe not only are different than previous calculations but are
also more reliable for reasons stated before.

\section*{Acknowledgements}
J.-U. Nabi would like to
acknowledge the support of the Higher Education Commission
Pakistan through project numbers  5557/KPK/NRPU/R$\&$D/HEC/2016 and  9-5(Ph-1-MG-7)Pak-Turk/R$\&$D/HEC/2017 and
Pakistan Science Foundation through project number
PSF-TUBITAK/KP-GIKI (02).

\end{document}